\documentclass[twocolumn]{aastex631}

\accepted{November 2, 2021}
\usepackage{amsfonts}
\usepackage{amsmath}
\usepackage{amssymb}
\usepackage{graphicx}
\usepackage{float}
\usepackage{csquotes}
\usepackage{physics}
\usepackage{xfrac}

\usepackage{mathrsfs}

\newcommand*\patchAmsMathEnvironmentForLineno[1]{%
  \expandafter\let\csname old#1\expandafter\endcsname\csname #1\endcsname
  \expandafter\let\csname oldend#1\expandafter\endcsname\csname end#1\endcsname
  \renewenvironment{#1}%
     {\linenomath\csname old#1\endcsname}%
     {\csname oldend#1\endcsname\endlinenomath}}% 
\newcommand*\patchBothAmsMathEnvironmentsForLineno[1]{%
  \patchAmsMathEnvironmentForLineno{#1}%
  \patchAmsMathEnvironmentForLineno{#1*}}%
\AtBeginDocument{%
\patchBothAmsMathEnvironmentsForLineno{gather}
}
\newcommand{\lb}{\left(}
\newcommand{\rb}{\right)}
\newcommand{\lsq}{\left[}
\newcommand{\rsq}{\right]}
\newcommand{\lc}{\left\{}
\newcommand{\rc}{\right\}}

\newcommand{\kB}{k_\text{B}}

\newcommand{\bbS}{\mathbb{S}}
\newcommand{\bbSx}{\bbS_\textup{X}}
\newcommand{\bbSxt}{\bbS_\textup{X}^+}
\newcommand{\bbSy}{\bbS_\textup{Y}}
\newcommand{\bbSyt}{\bbS_\textup{Y}^+}
\newcommand{\bbSz}{\bbS_\textup{Z}}
\newcommand{\bbSzt}{\bbS_\textup{Z}^+}

\newcommand{\varphie}{\varphi} %_\textup{e}}
\newcommand{\varthetae}{\vartheta} %_\textup{e}}
\newcommand{\psie}{\psi} %_\textup{e}}

\newcommand{\unitvector}{\begin{bmatrix} 0 \\ 0 \\ 1 \end{bmatrix}}

\DeclareMathOperator{\atantwo}{atan2}
\newcommand{\intty}{\int\limits_{-\infty}^{+\infty}}
\newcommand*\diff{\mathop{}\!\mathrm{d}}
\newcommand{\mf}{\mathbf}
\newcommand{\reducedspacemodifier}{\star}
\newcommand{\bs}{\boldsymbol}

\newcommand{\bbT}{\mathbb{T}}
\newcommand{\p}{\partial}

\newcounter{appendixcounter}

\newcommand{\myedit}[1]{#1}

\begin{document}

\title{Trajectory-based simulation of far-infrared CIA profiles of CH$_4-$N$_2$ for modeling Titan's atmosphere}

\correspondingauthor{Finenko A. A.}
\email{artem.finenko@cfa.harvard.edu, artfin@mail.ru}

\author[0000-0002-4730-8841]{Finenko A. Artem}
\affiliation{Center for Astrophysics \textbar Harvard \& Smithsonian,  Atomic and Molecular Physics Division, 60 Garden Street, Cambridge MA 02138, USA}
\affiliation{Department of Chemistry, Lomonosov Moscow State University, GSP-1, 1-3 Leninskiye Gory, Moscow 119991, Russia}

\author[0000-0002-5433-5661]{Bruno B{\'e}zard}
\affiliation{LESIA, Observatoire de Paris, Universit{\'e} PSL, CNRS, Sorbonne Universit{\'e}, Universit{\'e} de Paris, place Jules Janssen, 92195 Meudon, France}

\author[0000-0003-4763-2841]{Iouli E. Gordon}
\affiliation{Center for Astrophysics \textbar Harvard \& Smithsonian,  Atomic and Molecular Physics Division, 60 Garden Street, Cambridge MA 02138, USA}

\author{Daniil N. Chistikov}
\affiliation{Department of Chemistry, Lomonosov Moscow State University, GSP-1, 1-3 Leninskiye Gory, Moscow 119991, Russia}
\affiliation{Obukhov Institute of Atmospheric Physics, Russian Academy of Sciences, 3 Pyzhevsky per., Moscow 119017,  Russia}
\affiliation{Institute of Quantum Physics, Irkutsk National Research Technical University, 83 Lermontov str., Irkutsk 664074, Russia}

\author[0000-0001-6639-4936]{Sergei E. Lokshtanov}
\affiliation{Department of Chemistry, Lomonosov Moscow State University, GSP-1, 1-3 Leninskiye Gory, Moscow 119991, Russia}

\author{Sergey V. Petrov}
\affiliation{Department of Chemistry, Lomonosov Moscow State University, GSP-1, 1-3 Leninskiye Gory, Moscow 119991, Russia}

\author[0000-0002-4677-9692]{Andrey A. Vigasin}
\affiliation{Obukhov Institute of Atmospheric Physics, Russian Academy of Sciences, 3 Pyzhevsky per., Moscow 119017,  Russia}

\begin{abstract}
We report the results of the trajectory-based simulation of far-infrared collision-induced absorption (CIA) due to CH$_4-$N$_2$ pairs at temperatures between 70 and 400~K. Our analysis utilizes recently calculated high-level potential energy (PES) and induced dipole surfaces (IDS) [Finenko, A.~A., Chistikov, D.~N., Kalugina, Y.~N., Conway E.~K., Gordon, I.~E., Phys. Chem. Chem. Phys., 2021, doi: \href{http://doi.org/10.1039/d1cp02161c}{10.1039/d1cp02161c}]. 
Treating collision partners as rigid rotors, the time evolution of interaction-induced dipole is accumulated over a vast ensemble of classical trajectories and subsequently transformed into CIA spectrum via Fourier transform. In our calculations, both bound and unbound states are properly accounted for, and the rigorous theory of lower-order spectral moments is addressed to check the accuracy of simulated profiles.  
Classically derived trajectory-based profiles are subject to two approximate desymmetrization procedures so that resulting profiles conform to the quantum principle of detailed balance.
The simulated profiles are compared to laboratory measurements and employed for modeling Titan's spectra in the 50--500~cm$^{-1}$ range. Based on the desymmetrized simulated profiles, a new semi-empirical model for CH$_4-$N$_2$ CIA is proposed for modeling Titan's infrared spectra. Synthetic spectra derived using this model yield an excellent agreement with the data recorded by the Composite Infrared Spectrometer (CIRS) aboard the Cassini spacecraft at low and high emission angles.
\end{abstract}


%%%%%%%%%%%%%%%%%%%%%%%%%%%%%%%%%%%%%%%%%%%%%%%%%%%%%%%%%%%%%%%%%%%%%%%%%%%%%%%%%%%%%%%%%%%%%%%
\section{Introduction}
\label{sec:introduction}
%%%%%%%%%%%%%%%%%%%%%%%%%%%%%%%%%%%%%%%%%%%%%%%%%%%%%%%%%%%%%%%%%%%%%%%%%%%%%%%%%%%%%%%%%%%%%%%

Nitrogen and methane were well established as the primary atmospheric constituents of Titan's atmosphere after the Voyager 1 encounter \citep{Hunten1984}. With a concentration of roughly 95\%, nitrogen is the most abundant species in the atmosphere of Titan. Methane, while being less abundant, is crucial to the maintenance of the thick nitrogen atmosphere; the atmosphere would gradually condense in the absence of warming resulting from the far-infrared CH$_4-$N$_2$ collision-induced absorption and hydrocarbon haze \citep{Lorenz1997}. The photochemistry of nitrogen and methane leads to the formation of complex hydrocarbons and nitriles in the atmosphere of Titan. 

\begin{figure}
    \centering
    \includegraphics[width=\linewidth]{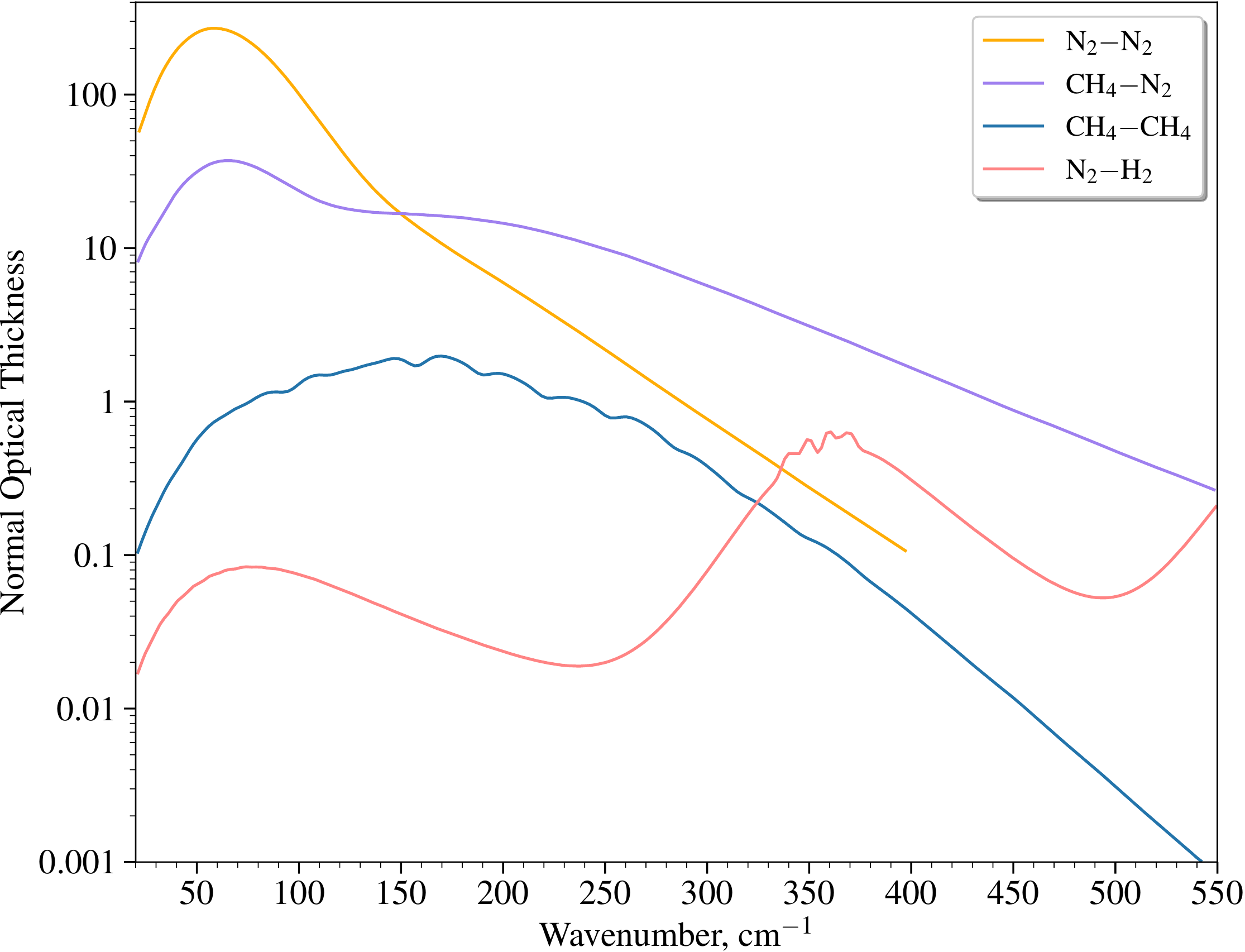}
    \caption{Normal optical thicknesses of Titan's atmosphere. Collision-induced intensities due to N$_2-$N$_2$ \citep{Chistikov2019}, CH$_4-$N$_2$ (current work), CH$_4-$CH$_4$ \citep{Borysow1987-CH4-CH4}, and N$_2-$H$_2$ \citep{Borysow1986-H2-N2} are shown separately. The contribution from CH$_4-$N$_2$ CIA is obtained employing semi-classical model as described in Section~\ref{sec:modeling-cassini-measurements} (see Eq.~\eqref{modeling:linear-model}). The calculations were performed utilizing atmosphere model of \citet{Bezard2020}.}
    \label{fig:optical-thickness}
\end{figure}

Figure~\ref{fig:optical-thickness} highlights the magnitude and wavenumber dependence of the major CIA contributions of relevant collision pairs in Titan's atmosphere.   
The maximum of spectral radiance from relatively cold Titan's surface falls in the vicinity of 200 cm$^{-1}$. The most abundant atmospheric species like N$_2$, CH$_4$, and H$_2$ have no permanent dipole and are therefore nearly transparent in the relevant far-infrared spectral range. As a result, the opacity of Titan's atmosphere is largely determined by continuum absorption dominated by the collision-induced component that belongs to molecular pairs formed from the main atmospheric constituents: N$_2–$N$_2$, CH$_4-$N$_2$, CH$_4–$CH$_4$, N$_2–$H$_2$, H$_2–$H$_2$, CH$_4–$H$_2$. 
Due to the large abundance of N$_2$ and relatively small rotational constant, $B \approx 2$ cm$^{-1}$, the N$_2-$N$_2$ CIA dominates the tropospheric opacity for wavenumbers shortward of $\sim150$ cm$^{-1}$, whereas the CH$_4-$N$_2$ CIA is the prevailing source of opacity in the 150--450 cm$^{-1}$ region \citep{Samuelson1997}. Several previous analyses \citep{Anderson2011, Tomasko2008, deKok2010} concluded that the opacity due to CH$_4-$N$_2$ could not be correctly reproduced using the model of \citet{Borysow1993}. \citet{Courtin1995} and \citet{Samuelson1997} suggested that missing opacity could be attributed to a large supersaturation of methane in low-latitude regions of the upper troposphere. 
However, measurements performed using Gas Chromatograph Mass Spectrometer (GCMS) on the Huygens probe indicated that there is no methane supersaturation in the lower atmosphere at 10$^\circ$S latitude \citep{Niemann2005}.   
Later studies \citep{Tomasko2008, deKok2010} suggested that \citet{Borysow1993}'s CIA data is possibly systematically in error at Titan's temperatures. Both studies derived that a multiplicative factor of approximately 1.5 needs to be applied to \citet{Borysow1993} data to obtain satisfactory fits of the Cassini's Composite Infrared Spectrometer (CIRS) observations in the 150--450~cm$^{-1}$ region. \citet{Anderson2011} proposed the heuristic correction factor of the same magnitude. \citet{Bezard2020} reported that the wavenumber-dependent correction of \citet{Anderson2011} produces too much absorption between 180 and 300 cm$^{-1}$. Instead, \citet{Bezard2020} chose to use a wavenumber-independent multiplicative factor, the value of which was refined to 1.52.  

The most accurate laboratory measurements of collision-induced absorption in CH$_4-$N$_2$ gaseous mixtures were performed up to 550~cm$^{-1}$ at 162~K and up to 670~cm$^{-1}$ at 195 and 297~K by \citet{Birnbaum1993}. An earlier effort by \citet{Dagg1986} provides experimentally measured spectra at 126, 149, 179, and 212~K over the frequency region of 40--400 cm$^{-1}$. These measurements, however, are affected by a substantial noise, most notably, at the edge of the covered frequency range at 350--400~cm$^{-1}$.  
The temperature of Titan's atmosphere varies from 70 to 94~K, close to the surface, to 120~K in the upper atmosphere. Thus, the temperatures at which the absorption spectra are required for modeling Titan's atmosphere are much lower than those at which laboratory experiments were carried out. 

Despite significant efforts undertaken in the past to characterize the CIA sources of opacity for Titan, the required data remain either incomplete or not accurate enough to be used in the modeling of the atmospheric radiation. A series of papers by Borysow and Frommhold \citep{Borysow1986-H2-N2, Borysow1986-N2-N2, Borysow1986-CH4-H2, Borysow1987-CH4-CH4, Borysow1993} provided a database of calculated opacities based on the effective isotropic potentials combined with adjusted short-range induced dipoles. The intrinsic parameters of the induced dipole were derived through the fitting to experimental spectra. This approach inherently depends on the quality of experimental data since the possible systematic errors in the fitted data would be introduced in the model parameters. 
Furthermore, the validity of such profiles modeled far outside the temperature range for which the intrinsic parameters have been fitted is under question.   
In the case of CH$_4-$N$_2$, \citet{Borysow1993} opted to perform fitting to measurements by \citet{Birnbaum1993}.  

Recently, sophisticated first-principles approaches were developed and applied to simulate far-infrared absorption in N$_2-$N$_2$ using quantum calculation with proper account for intermolecular anisotropy \citep{Karman2015}, molecular dynamics \citep{Hartmann2011-II}, and classical trajectory-based simulation \citep{Chistikov2019}. Encouraged by the success of the trajectory-based approach in the case of N$_2-$N$_2$ \citep{Chistikov2019, HITRAN2020} and CO$_2-$Ar \citep{Chistikov2021}, we extended it towards linear molecules interacting with spherical top molecules. Our consideration relies on the \textit{ab initio} potential energy and induced dipole surfaces for CH$_4-$N$_2$ calculated at CCSD(T)-F12/\text{aug}-cc-pVTZ and CCSD(T)/\text{aug}-cc-pVTZ levels of theory, respectively, reported in \citet{Finenko2021}. Section~\ref{sec:cia-methodology} presents the details of our trajectory-based simulation of CIA in CH$_4-$N$_2$ molecular pairs, including transitions characteristic to unbound (free) and quasi-bound molecular pairs as well as to true bound dimers. Section~\ref{sec:modeling-cassini-measurements} shows that the semi-empirical model based on the \textit{ab initio} simulated CH$_4-$N$_2$ CIA profiles allowed us to reproduce Titan's spectra between 50 and 650~cm$^{-1}$ recorded by the Composite Infrared Spectrometer (CIRS) aboard the Cassini spacecraft with unprecedented accuracy. Potential implications of the simulated profiles for greenhouse warming of early Earth's atmosphere are discussed in Section~\ref{sec:archean-earth}. 

%%%%%%%%%%%%%%%%%%%%%%%%%%%%%%%%%%%%%%%%%%%%%%%%%%%%%%%%%%%%%%%%%%%%%%%%%%%%%%%%%%%%%%%%%%%%%%%
\section{Methodology of CIA spectrum calculation}
\label{sec:cia-methodology}
%%%%%%%%%%%%%%%%%%%%%%%%%%%%%%%%%%%%%%%%%%%%%%%%%%%%%%%%%%%%%%%%%%%%%%%%%%%%%%%%%%%%%%%%%%%%%%%

%%%%%%%%%%%%%%%%%%%%%%%%%%%%%%%%%%%%%%%%%%%%%%%%%%%%%%%%%%%%%%%%%%%%%%%%%%%%%%%%%%%%%%%%%%%%%%%
\subsection{Summary}
\label{subsec:summary}
%%%%%%%%%%%%%%%%%%%%%%%%%%%%%%%%%%%%%%%%%%%%%%%%%%%%%%%%%%%%%%%%%%%%%%%%%%%%%%%%%%%%%%%%%%%%%%%

In first-order perturbation theory, the binary absorption coefficient $\alpha(\nu, T)$ in a mixture of gases can be represented as \citep{Frommhold2006}
\begin{gather}
    \begin{aligned}
        \alpha(\nu, T) = &\frac{\tau \lb \nu, T \rb}{\rho_1 \rho_2} = \\
        &\frac{\lb 2 \pi \rb^4 N_L^2}{3 h c} \nu \lb 1 - \exp \lb -\beta h c \nu \rb \rb V G(\nu, T),
    \end{aligned}
    \label{general:alpha}
\end{gather}
where 
\begin{gather}
    \tau(\nu, T) = L^{-1} \ln \lb I_0 / I \rb
\end{gather}
is the absorption coefficient, $\rho_1$ and $\rho_2$ are the number densities of gas constituents; $\beta = \lb \kB T \rb^{-1}$ is the reciprocal temperature in units of energy,
$k_B$, $h$, $c$, and $N_L$ are the Boltzmann and Planck constants, speed of light, and Loschmidt number, respectively, $V$ designates the gas volume, and $\nu$ is the wavenumber in the reciprocal centimeters.
The spectral function $G(\nu, T)$ is given by the Fourier transform of the time correlation function (TCF) $C(t, T)$
\begin{gather}
    G(\nu, T) = \frac{1}{2 \pi} \intty C(t, T) e^{-2 \pi i c \nu t} \diff{t},
    \label{general:spectral-function}
\end{gather}
which is defined as the statistically-averaged product of the time-shifted dipoles relevant to the compound molecular pair
\begin{gather}
    C(t, T) = \frac{V}{4 \pi \varepsilon_0} \langle \boldsymbol{\mu}(0) \cdot \boldsymbol{\mu}(t) \rangle,
    \label{general:correlation-function}
\end{gather}
where $\varepsilon_0$ is the vacuum permittivity, and angular brackets denote phase-space averaging. 
The spectral function conforms to the so-called detailed balance condition \citep{Frommhold2006}
\begin{gather}
    G(-\nu, T) = e^{-\beta h c \nu} G(\nu, T)
    \label{general:detailed-balance}
\end{gather}
causing the striking asymmetry of the observed spectral profiles.
If exact quantum formalism is employed to simulate the spectral function, the detailed balance condition \eqref{general:detailed-balance} is satisfied naturally. However, if one chooses to address the approach utilizing classical mechanics, the resulting spectral function is symmetric
\begin{gather}
    G_\text{cl}(-\nu, T) = G_\text{cl}(\nu, T),
    \label{general:classical-detailed-balance}
\end{gather}
in violation of Eq.~\eqref{general:detailed-balance}. \myedit{This is caused by the time-reversal invariance of classical dynamics, which induces the symmetry of the corresponding TCF.} An appealing approach to approximately correct this major defect is through the use of the so-called desymmetrization procedures \citep{Frommhold2006, Borysow1985}. These are \textit{ad hoc} prescriptions for constructing spectral functions that obey detailed balance condition from the classical counterpart. A variety of such prescriptions were previously proposed to reproduce profiles rendered from the quantum approach for model problems \citep{Schofield1960, Berens1981, Bader1994, Egelstaff1962}.  
Let us formulate several known prescriptions in terms of the quantum correction factor $\mathcal{F}(\nu, T)$ defined by     
\begin{gather}
    G(\nu, T) = \mathcal{F}(\nu, T) G_\text{cl}(\nu, T).
    \label{general:quantum-correction}
\end{gather}
Schofield's procedure \citep{Schofield1960}
\begin{gather}
    \mathcal{F}_\text{Sch}(\nu, T) = e^{\beta h c \nu / 2}
    \label{general:D3-correction}
\end{gather}
was demonstrated to provide a satisfactory semi-classical correction to classically simulated CIA rototranslational bands for multiple collisional complexes \citep{Hartmann2011-II, Bussery-Honvalut2014, Chistikov2019, Odintsova2021}.
\myedit{The more sophisticated quantum correction factor due to \citet{Egelstaff1962} was shown to provide the most accurate description of spectra in comparison to other corrections, at least in the case of freely rotating linear molecules \citep{Borysow1994}. The extension of Egelstaff's procedure to the case of interacting polyatomics is, therefore, somewhat more grounded than any other correction suggested in the literature.} \citet{Frommhold2006} suggested a slight variant of Egelstaff's procedure enhanced by the constant rescaling factor. This variant given by 
\begin{widetext}
\begin{gather}
    \mathcal{F}_\text{Fromm}(\nu, T) = \frac{e^{\beta h c \nu / 2}}{G_\text{cl}(\nu, T)} \frac{C_\text{cl}(0)}{C_\text{cl}(\beta h c \nu / \sqrt{2})}
    \int\limits_{-\infty}^{+\infty} C_\text{cl} \lb \sqrt{t^2 + \lb \beta \hbar / 2 \rb^2}, T \rb e^{-2 \pi i c \nu t} \diff{t}
    \label{general:D4a-correction}
\end{gather}
\end{widetext}
provides the closest approximation to the quantal spectral profiles for the purely translational spectra \citep{Frommhold2006}.
As a final result of our consideration, we adopt the semi-classical binary absorption coefficient
\begin{gather}
    \alpha_\text{SC}(\nu, T) = \frac{\lb 2 \pi \rb^4 N_L^2}{3 h c} \nu \lb 1 - \exp \lb - \beta h c \nu \rb \rb \mathcal{F}(\nu, T) V G_\text{cl}(\nu, T),
    \label{general:SC-alpha}
\end{gather}
where quantum correction factor $\mathcal{F}(\nu, T)$ is represented either by Schofield's \eqref{general:D3-correction} or Frommhold's procedure \eqref{general:D4a-correction}.

\subsection{Trajectory-based approach}

Invoking Hamilton formalism, let us introduce the generalized coordinates describing the position and orientation of moieties relative to the space-fixed frame placed at the center of mass of the bimolecular complex. Let $\mf{R} =  \lb R, \Phi, \Theta \rb$ be the vector connecting the monomers' centers of mass. The Euler angles $\zeta_1 = \lb \varphi, \vartheta, \psi \rb$ and $\zeta_2 = \lb \eta, \chi \rb$ describe an active rotation of CH$_4$ and N$_2$ moiety, respectively, from an initial position, in which a reference frame fixed on moiety is parallel to the space-fixed frame, to its present position.

In the space-fixed frame of reference, the Hamiltonian of the CH$_4-$N$_2$ system breaks up into
\begin{gather}
    H = K_\text{tr} + K_{\text{rot}_1} + K_{\text{rot}_2} + U,
    \label{approach:hamiltonian}
\end{gather}
where $U$ is potential energy function and the kinetic energies of translational and rotational motions are given by   
\begin{gather}
    \begin{aligned}
        K_\text{tr} &= \frac{p_R^2}{2 \mu} + \frac{p_\Theta^2}{2 \mu R^2} + \frac{p_\Phi^2}{2 \mu R^2 \sin^2 \Theta}, \\
        K_{\text{rot}_1} &= \frac{\lb p_{\varphi} - p_{\psi} \cos \vartheta \rb^2}{2 I_1 \sin^2 \vartheta} + \frac{p_\psi^2 + p_\vartheta^2}{2 I_1}, \\
        K_{\text{rot}_2} &= \frac{p_{\chi}^2}{2 I_2} + \frac{p_{\eta}^2}{2 I_2 \sin^2 \chi}.
    \end{aligned}
    \label{approach:hamiltonian-terms}
\end{gather}
Here $\mu$, $I_1$, and $I_2$ denote the reduced mass of the molecular pair and moments of inertia of CH$_4$ and N$_2$, respectively. The momentum conjugated to the generalized variable $\xi$ is designated via $p_\xi$. Hereafter, the generalized coordinates and conjugated momenta are assembled into vectors $\mf{q}$ and $\mf{p}$, respectively.  

\begin{figure}
    \centering
    \includegraphics[width=\linewidth]{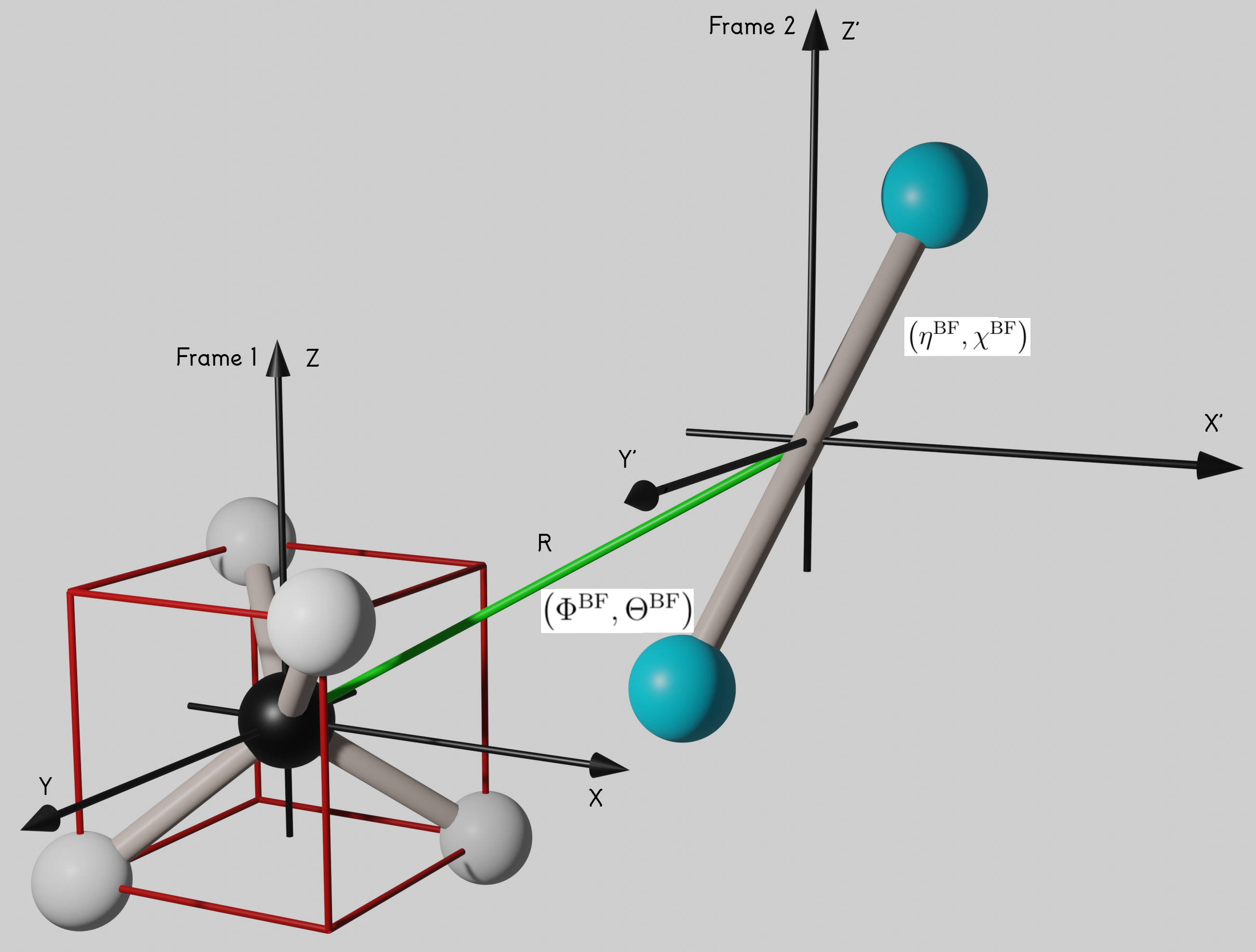}
    \caption{Body-fixed coordinate frame of the CH$_4-$N$_2$ complex.}
    \label{fig:coordinate-frame}
\end{figure}

We consider the time evolution of the bimolecular complex in terms of the solution of the Hamilton equations arising from the Hamiltonian \eqref{approach:hamiltonian}:
\begin{gather}
    \begin{aligned}
        \dot{R} &= \frac{p_R}{\mu}, \\
        \dot{p}_R &= \frac{p_\Theta^2}{\mu R^3} + \frac{p_\Phi^2}{\mu R^3 \sin^2 \Theta} - \frac{\partial U}{\partial R}, \\
        \dot{\Phi} &= \frac{p_\Phi}{\mu R^2 \sin^2 \Theta}, \\
        \dot{p}_\Phi &= -\frac{\partial U}{\partial \Phi}, \\
        \dot{\Theta} &= \frac{p_\Theta}{\mu R^2}, \\
        \dot{p}_\Theta &= \frac{p_\Phi^2 \cos \Theta}{\mu R^2 \sin^3 \Theta} - \frac{\partial U}{\partial \Theta}, \\
        \dot{\varphi} &= \frac{p_{\varphi} - p_{\psi} \cos \vartheta}{I_1 \sin^2 \vartheta}, \\
        \dot{p}_{\varphi} &= -\frac{\partial U}{\partial \varphi}, \\
        \dot{\vartheta} &= \frac{p_\vartheta}{I_1}, \\
        \dot{p}_\vartheta &= \frac{\lb p_\varphi - p_{\psi} \cos \vartheta \rb \lb p_\varphi \cos \vartheta - p_\psi \rb}{I_1 \sin^3 \vartheta} - \frac{\partial U}{\partial \vartheta}, \\
        \dot{\psi} &= \frac{p_\psi - p_\varphi \cos \vartheta}{I_1 \sin^2 \vartheta}, \\
        \dot{p}_\psi &= -\frac{\partial U}{\partial \psi}, \\
        \dot{\eta} &= \frac{p_\eta}{I_2 \sin^2 \chi}, \\
        \dot{p}_\eta &= -\frac{\partial U}{\partial \eta}, \\
        \dot{\chi} &= \frac{p_\chi}{I_2}, \\
        \dot{p}_\chi &= \frac{p_\eta^2 \cos \chi}{I_2 \sin^3 \chi} - \frac{\partial U}{\partial \chi}.
    \end{aligned}
    \label{general:dynamical-equations}
\end{gather}
The potential energy and induced dipole surfaces are commonly expanded in the body-fixed frame in order to reduce the number of angular degrees of freedom. Here we utilized the PES and IDS computed at a high level of \textit{ab initio} theory in \citet{Finenko2021}. Those were constructed in the frame in which the orientation of the CH$_4$ is fixed in its initial orientation. The schematic representation of the frame is given in Figure~\ref{fig:coordinate-frame}. The origin is placed on the carbon atom of the CH$_4$ monomer. The orientation of the vector $\mf{R}$ is described by angles $\lb \Phi^\text{BF}, \Theta^\text{BF} \rb$. Frame 2, parallel to Frame 1 fixed to CH$_4$ molecule, has its origin in the center of mass of N$_2$. Orientation of N$_2$ molecule with respect to Frame 2 is described by angles $\lb \eta^\text{BF}, \chi^\text{BF} \rb$. 
The main computational impediment that we face when solving dynamical equations \eqref{general:dynamical-equations} is to find the derivatives of the potential energy $U$ with respect to the space-fixed angles $\Phi, \Theta, \varphi, \vartheta, \psi, \eta, \chi$.
The numerical scheme of calculating the desired derivatives is described in Appendix~\ref{appendix:transformation-of-angles}. 

From a formal point of view, one can opt for direct simulation of the TCF using classical trajectories method for the purpose of then transforming it to spectral function according to Eq.~\eqref{general:spectral-function}. 
The Monte Carlo approach provides means for the simulation of a TCF since the high dimensionality of a phase space forbids practical use of deterministic numerical integration schemes. The convergence issues at large frequencies $\nu$, however, hinder the application of the direct approach for free and quasi-bound states. An alternative integral representation of the spectral function is often adapted in the classical trajectories modeling of these pair states \citep{Kranendonk1973, Oparin2017, Chistikov2019}. In the seminal work by \citet{Kranendonk1973}, the transformation into an integral expression over complete trajectories is suggested, alleviating the convergence difficulties. \citet{Chistikov2019} suggested a similar integral representation suitable for a canonical set of Hamilton variables. Adopting the latter integral representation for free \myedit{(F)} and quasi-bound \myedit{(Q)} states, the corresponding classical spectral function $G_\text{cl}^\text{F+Q}$ can be written as 
\begin{widetext}
    \vspace*{-0.5cm}
    \begin{gather}
        G_\text{cl}^\text{F+Q}(\nu, T) \approx \frac{1}{2 \pi Q(T)} \int\limits_{-\infty}^{0} \frac{-p_R}{\mu} \diff{p_R} \idotsint\limits_{H > 0} \exp \lb -\beta H \rb \diff{ \mf{q}^\reducedspacemodifier} \diff{\mf{p}^\reducedspacemodifier} 
    \Bigg\vert \intty \bs{\mu}(t) e^{-2\pi i c \nu t} \diff{t} \Bigg\vert^2.
        \label{general:prmu-spectral-function}
    \end{gather}
\end{widetext}
Here, $Q(T)$ represents the partition sum
\begin{gather}
    Q(T) = \int e^{-\beta H} \diff{\mf{q}} \diff{\mf{p}}.
    \label{general:partition-sum}
\end{gather}
An instantaneous value of the dipole moment vector of the collisional complex is designated with $\boldsymbol{\mu}(t)$, and vectors $\mf{q}^\reducedspacemodifier$ and $\mf{p}^\reducedspacemodifier$ are obtained upon exclusion of the coordinate $R$ and conjugated momentum $p_R$ from the vectors $\mf{q}$ and $\mf{p}$, respectively. 

Integral representation \eqref{general:prmu-spectral-function} should be understood in the following way. First, some fixed value of intermolecular distance $R_\text{max}$ is chosen, and a hyperplane in the phase space corresponding to the chosen value is drawn. All trajectories passing through the selected hyperplane are properly accounted for in the representation \eqref{general:prmu-spectral-function}. There are infinitely many far fly-by trajectories for which the intermolecular distance $R$ always exceeds $R_\text{max}$ that are naturally excluded from the expression \eqref{general:prmu-spectral-function}. The larger value of $R_\text{max}$ is chosen, the more negligible contribution to the spectral function gets from the excluded trajectories since the induced dipole vanishes at extreme separation.
As discussed in \citet{Chistikov2021}, there is a family of quasi-periodic trajectories that, while having positive energy, do not result in the spontaneous breakdown of the molecular complex into monomers within two-body consideration. The time evolutions of dipole moment along those peculiar trajectories contribute to the spectral function $G_\text{cl}^\text{F+Q}$. The sampling of quasi-periodic trajectories through representation \eqref{general:prmu-spectral-function} can be achieved only when the value of $R_\text{max}$ is somewhat close to the potential well. Thus it becomes clear that no value of $R_\text{max}$ allows for taking into account contributions from all significant trajectories.   
The spectral signatures from the quasi-periodic trajectories are restricted to comparatively low frequencies and hence are adequately simulated through the calculation of the TCF. Selecting a large value of $R_\text{max}$ for the representation \eqref{general:prmu-spectral-function} permits simulation of the spectral profile, excluding the quasi-periodic contributions. 
\myedit{Therefore we opt to combine a TCF-based spectral function at lower frequencies with the one based on the representation~\eqref{general:prmu-spectral-function} in the spectral wing using a basic stitching procedure (see  Section~\ref{subsec:stitching-procedure}). Contributions from all the mentioned pair states are sufficiently accounted for in the combined spectral function, which is confirmed by the agreement between spectral function moments and their phase-space counterparts described in Section~\ref{sec:spectral-moments}.}
\myedit{The calculation of the bound states' spectral function is addressed through the calculation of the TCF supplemented by the proper averaging over the true bound domain in the phase space as described below.}

To numerically evaluate the multivariate integral \eqref{general:prmu-spectral-function}, we invoke the importance sampling approach with proposal distribution $\exp \lb -\beta H \rb$ over the domain of unbound states. The sampling algorithm is outlined in Appendix~\ref{appendix:initial-condition-generation}. Given a supply of $N$ random variables $\lc \mf{q}_k^\reducedspacemodifier, p_{R, k}, \mf{p}_k^\reducedspacemodifier \rc$, the Monte Carlo estimate is given by (for a more detailed discussion see \citet{Chistikov2019})
\begin{widetext}
    \vspace*{-0.5cm}
    \begin{gather}
        G_\text{cl}^\text{F+Q}(\nu, T) \approx \frac{3}{4 \pi R_\text{max}} \frac{1}{N} \sum_{k = 1}^{N} \frac{-p_{R, k}}{\mu} \Bigg\vert \intty \bs{\mu} \lb t; R_\text{max}, \mf{q}_k^\reducedspacemodifier, p_{R, k}, \mf{p}_k^\reducedspacemodifier \rb e^{-2 \pi i c \nu t} \diff{t} \Bigg\vert^2.
        \label{general:prmu-spectral-function-monte-carlo}
    \end{gather}
\end{widetext}
Here, $\boldsymbol{\mu} \lb t; \cdot \rb$ is the value of dipole moment at the time $t$ during classical evolution of trajectory starting initially at the given phase point.

The Monte Carlo simulation of the TCF is addressed by calculating an ensemble of classical trajectories emerging from Boltzmann-distributed initial points. The sampling from the Boltzmann \myedit{distribution} is performed in the course of a two-step rejection-based procedure summarized in Appendix~\ref{appendix:initial-condition-generation}. Given $N$ phase space points $\lc \mf{q}_k, \mf{p}_k \rc$ supplied by the rejection algorithm, the TCF in the domain $\Omega = \lc H < 0 \rc$ or $\lc H > 0 \rc$ is obtained according to
\begin{widetext}
    \vspace*{-0.5cm}
    \begin{gather}
        \begin{aligned}
        C_\Omega(t; T) = \frac{1}{4 \pi \varepsilon_0} \frac{V}{N} \frac{Q_\Omega \lb R_\text{min}, R_\text{max}; T \rb}{Q \lb T \rb} \sum_{i = 1}^{N} \frac{1}{2} \Big[ \boldsymbol{\mu} \lb 0; \mf{q}_i, \mf{p}_i \rb \boldsymbol{\mu} \lb t; \mf{q}_i, \mf{p}_i \rb  + \boldsymbol{\mu} \lb 0; \mf{q}_i, \mf{p}_i \rb \boldsymbol{\mu} \lb -t; \mf{q}_i, \mf{p}_i \rb \Big],
        \end{aligned}
    \label{general:correlation-function-monte-carlo}
    \end{gather}
\end{widetext}
where $Q_\Omega \lb R_\text{min}, R_\text{max}; T \rb$ is the partial partition sum defined by
\begin{widetext}
    \vspace*{-0.5cm}
    \begin{gather}
        Q_\Omega \lb R_1, R_2; T \rb = \int\limits_{R_1}^{R_2} \diff{R} \int\limits_{-\infty}^{+\infty} \diff{p_R} \idotsint\limits_{\Omega} \exp \lb -\frac{H}{k_\text{B} T} \rb \diff{\mf{q}^\star} \diff{\mf{p}^\star}.
        \label{general:partial-partition-function}
    \end{gather}
\end{widetext}
\myedit{Averaging of the products of dipole moments in Eq.~\eqref{general:correlation-function-monte-carlo}, containing either positive or negative time arguments, ensures time-reversibility of the TCF estimate: $C_\Omega(t; T) = C_\Omega(-t; T)$.}
In the case of rigid linear molecule-spherical top complex, the total partition sum can be expressed via the molecular complex parameters (see \citet{Chistikov2018} for related discussion)
\begin{gather}
    Q(T) = 32 \pi^3 V \lb 2 \pi k_\text{B} T \rb^4 \mu^{3/2} I_1^{3/2} I_2.
    \label{general:partition-function-analytical}
\end{gather}
The adaptive Monte Carlo integration method \texttt{VEGAS}\footnote{A C++11 template library for Monte Carlo integration: \url{https://github.com/cschwan/hep-mc}} \citep{Lepage1978} is utilized to compute the partial partition sum $Q_\Omega \lb R_\text{min}, R_\text{max}; T \rb$ prior to the calculation of the TCF. We opted to use $R_\text{min} = 4.0$ $a_0$ and $R_\text{max} = 35.0$ $a_0$ for the calculations described above.

The dipole variations along individual trajectories in Eq.~\eqref{general:correlation-function-monte-carlo} can be re-weighted to produce correlation function estimates at various temperatures. This computational technique is based on the premise that an individual classical trajectory is a microcanonical object that does not depend on temperature. The Monte Carlo staged multitemperature algorithm is described in detail in Section IV.C of \citet{Chistikov2021} and used here without modifications for the temperature range 70--400~K with a 10~K step.  

The correlation functions along individual trajectories were sampled at fixed intervals of \myedit{4.8 fs} until \myedit{$t_\text{max} = 1.58$~ns} is reached. Differential equation solver CVODE from the SUNDIALS suite \citep{SUNDIALS} was employed to propagate equations of motions~\eqref{general:dynamical-equations}. The obtained TCFs approach the positive limit as time goes to infinity. Because of this, one should take precautions before transforming TCFs into spectral functions according to Eq.~\eqref{general:spectral-function}. Direct application of standard numerical procedures implementing Fourier transform would cause artifacts to appear. The utilized processing technique allowing us to obtain proper spectral functions from the corresponding TCFs is explained in \citet{Chistikov2021}. Obtained spectral functions are subjected to smoothing using local quadratic fits with the variable-width window to remove the residual Monte Carlo noise. 
Subsequently, absorption coefficients are produced according to Eq.~\eqref{general:SC-alpha} with properly chosen quantum correction factors.

%%%%%%%%%%%%%%%%%%%%%%%%%%%%%%%%%%%%%%%%%%%%%%%%%%%%%%%%%%%%%%%%%%%%%%%%%%%%%%%%%%%%%%%%%%%%%%%
\subsection{Stitching procedure}
\label{subsec:stitching-procedure}
%%%%%%%%%%%%%%%%%%%%%%%%%%%%%%%%%%%%%%%%%%%%%%%%%%%%%%%%%%%%%%%%%%%%%%%%%%%%%%%%%%%%%%%%%%%%%%%

\begin{figure}
    \centering
    \includegraphics[width=\linewidth]{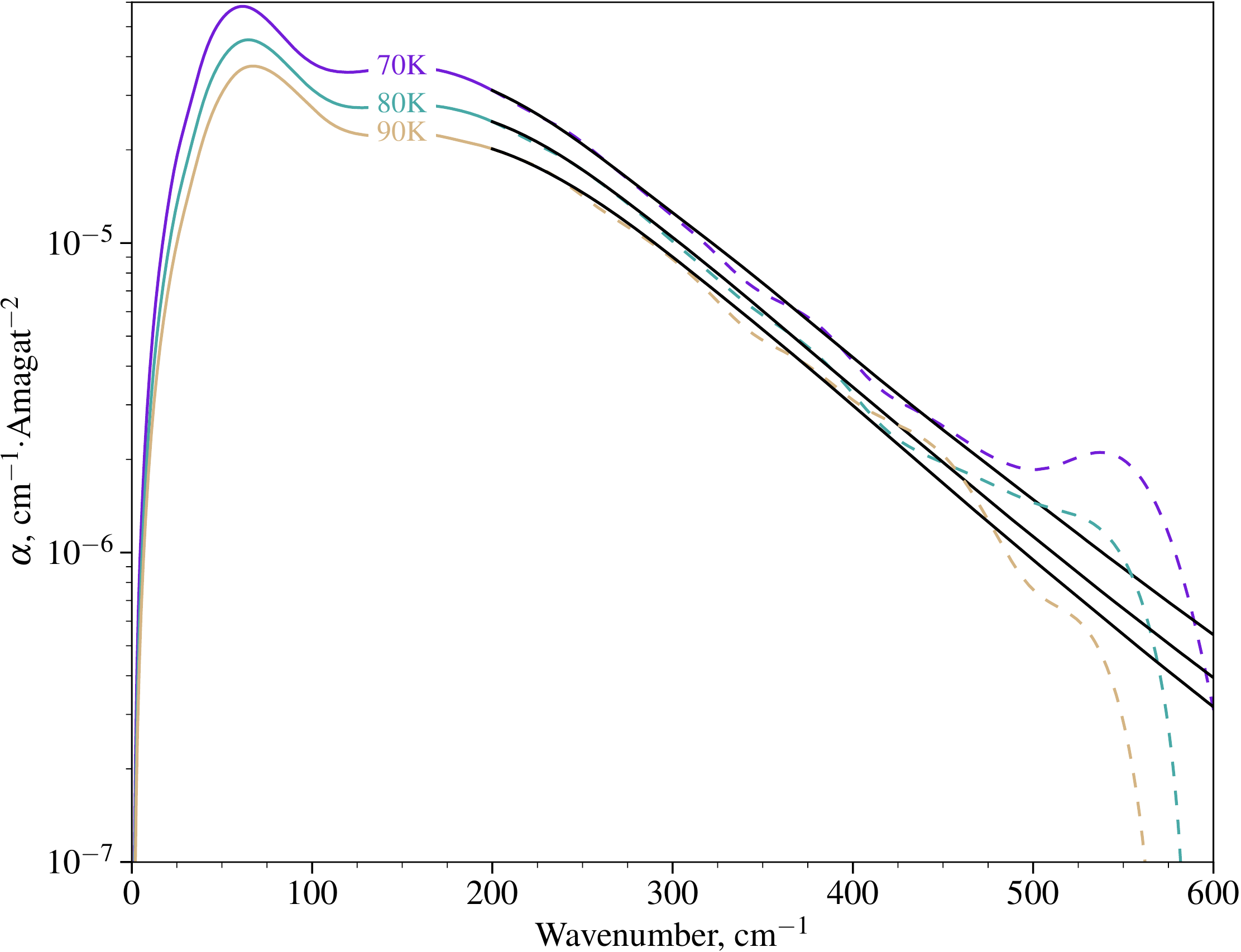}
    \caption{CH$_4-$N$_2$ CIA spectral profiles corresponding to unbound states obtained through TCFs processing (in color) and through formalism~\eqref{general:prmu-spectral-function} (in black) at 70--90~K. Both sets of profiles utilize Schofield's quantum correction factor. The TCF-based profiles longward of 200~cm$^{-1}$ are shown with dashed lines. Resultant profiles are obtained upon the stitching of two sets at 200~cm$^{-1}$.}
    \label{fig:stitching}
\end{figure}

\begin{figure*}
    \centering
    \includegraphics[width=0.49\linewidth]{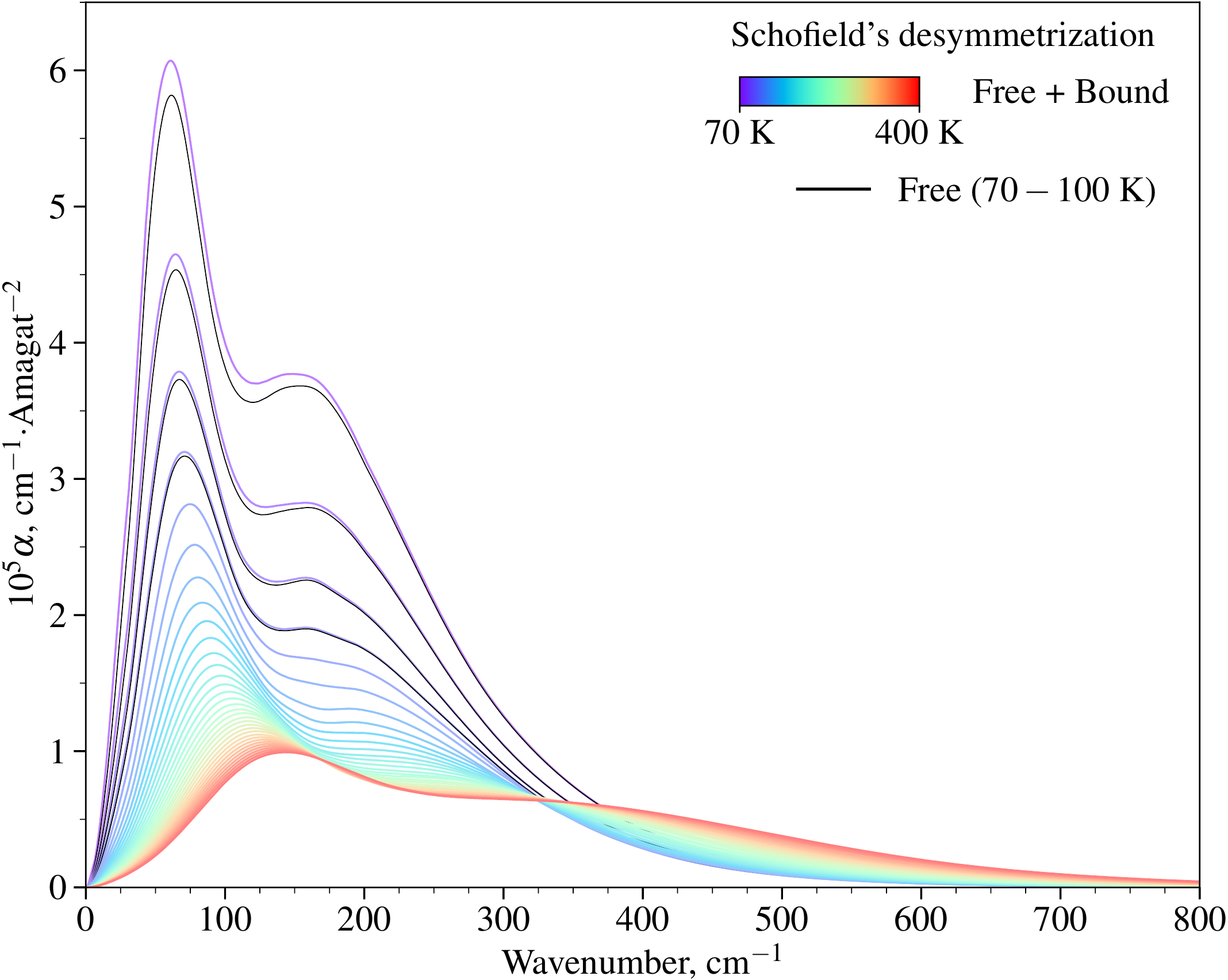}
    \includegraphics[width=0.49\linewidth]{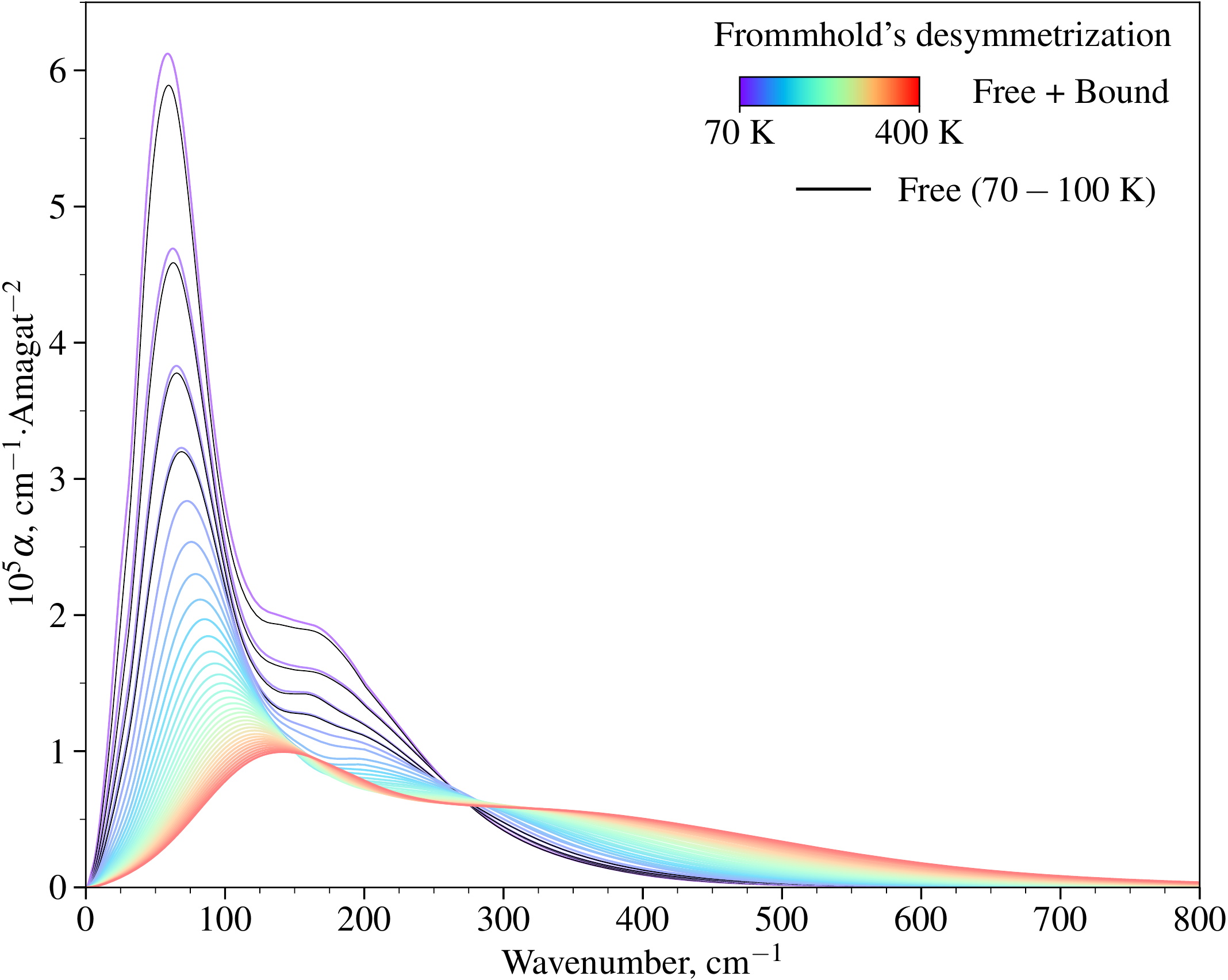}
    \caption{CH$_4-$N$_2$ CIA spectral profiles corresponding to both bound and unbound states (in color) from 70 to 400~K with a step of 10~K. The black lines represent profiles corresponding to unbound states at 70--100~K. The left panel shows profiles subjected to Schofield's quantum correction factor, whereas Frommhold's correction is utilized for the profiles on the right panel.}
    \label{fig:overview-multitemperature-spectra}
\end{figure*}

As mentioned above, the wings of the spectral profiles derived from the TCFs corresponding to unbound states exhibit \myedit{broad unphysical structure}, especially for low temperatures, demonstrated in Figure~\ref{fig:stitching}. \myedit{This structure fades slowly with an increasing number of trajectories and thus can be considered presumably as Monte Carlo noise.} To mitigate this effect, the wings of the spectral profiles derived through the formalism~(\ref{general:prmu-spectral-function}-\ref{general:prmu-spectral-function-monte-carlo}) are stitched continuously to TCF-based profiles at $\nu = 200$~cm$^{-1}$. The stitching is performed independently for the spectral profiles obtained using Schofield's and Frommhold's quantum correction factors. Thus obtained arrays of spectral profiles for both quantum corrections are shown in Figure~\ref{fig:overview-multitemperature-spectra}. 
The overall profile can be roughly decomposed into three spectral contributions having different dipole origins \citep{Birnbaum1993,Borysow1993}. First, absorption due to the dipole in the nitrogen molecule induced by the electric multipole field of methane manifests itself in the form of the pronounced peak. Second, spectral components due to the electric fields of nitrogen polarizing the methane molecule result in the formation of the spectral shoulder. Lastly, there are double-transition terms, usually much weaker, originating from the interaction of the gradient of the multipolar electric field of one molecule with the dipole-multipole polarizability tensor of another molecule.   
The intensity of the spectral shoulder greatly depends 
on the choice of the quantum correction factor, especially at low temperatures. Due to the rescaling factor introduced by \citet{Frommhold2006} to modify Egelstaff's desymmetrization, the conspicuous rotational peak caused by the collision-induced rotational transitions in N$_2$ molecule has an almost identical intensity for Schofield's and Frommhold's quantum corrections. 
It should be noted that the true bound states' contribution can be seen in Figure~\ref{fig:overview-multitemperature-spectra} as the difference between spectral profiles corresponding to both bound and unbound states and unbound states only. It is marginal for the lowest temperatures, amounting to 4\% of integral intensity at 70~K, and becoming effectively zero at temperatures higher than 120~K.  

%%%%%%%%%%%%%%%%%%%%%%%%%%%%%%%%%%%%%%%%%%%%%%%%%%%%%%%%%%%%%%%%%%%%%%%%%%%%%%%%%%%%%%%%%%%%%%%
\subsection{Spectral moments as convergence control parameters}
\label{sec:spectral-moments}
%%%%%%%%%%%%%%%%%%%%%%%%%%%%%%%%%%%%%%%%%%%%%%%%%%%%%%%%%%%%%%%%%%%%%%%%%%%%%%%%%%%%%%%%%%%%%%%

The convergence of obtained spectral profiles~\eqref{general:correlation-function-monte-carlo} with respect to the number of trajectories is controlled through the calculation of the lower-order spectral moments. The usefulness of spectral moments in this context arises from the fact that they can be derived from two significantly different prerequisites. On the one hand, the classical spectral moments can be expressed as the weighted integrals of the binary absorption coefficient
\begin{gather}
    M_n = 2 \int\limits_0^\infty \nu^{n - 1} \lsq 1 - \exp \lb -\frac{h c \nu}{k_\text{B} T} \rb \rsq^{-1} \alpha \lb \nu, T \rb \diff{\nu}.
    \label{general:spectral-moments-alpha}
\end{gather}
On the other hand, the lower-order spectral moments can be obtained through phase-space averaging according to \myedit{(see, \textit{e.g.}, \citet{Frommhold2006})}
\begin{gather}
    \begin{aligned}
        M_0 &= \frac{\lb 2 \pi \rb^4 N_L^2}{3 h} \frac{V}{4 \pi \varepsilon_0} \frac{1}{2 \pi c} \langle \boldsymbol{\mu}\myedit{(0)}^2 \rangle, \\
        M_2 &= \frac{\lb 2 \pi \rb^4 N_L^2}{3 h} \frac{V}{4 \pi \varepsilon_0} \frac{1}{\lb 2 \pi c \rb^3} \langle \dot{\boldsymbol{\mu}}\myedit{(0)}^2 \rangle.
    \end{aligned}
    \label{general:spectral-moments-phase-space}
\end{gather}
\myedit{When comparing spectral moments obtained through phase-space integration with the values derived from experimental spectra, one should take into account that the corresponding spectra obey different detailed balance conditions shown in Eqs.~\eqref{general:detailed-balance} and \eqref{general:classical-detailed-balance} (the issue is expounded upon in \citet{Finenko2021} and \citet{Chistikov2021}). To compare the moments issued from either the observations or the calculations, one must first introduce the respective correction. We chose to compare the moments that are derived utilizing classical detailed balance, Eq.~\eqref{general:classical-detailed-balance}. Consequently, prior to computing the moments from the laboratory-measured spectra using Eq.~\eqref{general:spectral-moments-alpha} a symmetrization procedure should be applied to produce the spectral shape conforming to the classical detailed balance. The symmetrization procedure is procured as an inverse to the desymmetrization procedure. For its simplicity, we opted to invert Schofield’s procedure so that the measured binary absorption coefficient is multiplied by the factor 
\begin{gather}
    \mathcal{F}_\text{inv-Sch} \lb \nu, T \rb = e^{-\beta h c \nu / 2}.
\end{gather}
}

\begin{figure*}
    \centering
    \includegraphics[width=0.49\linewidth]{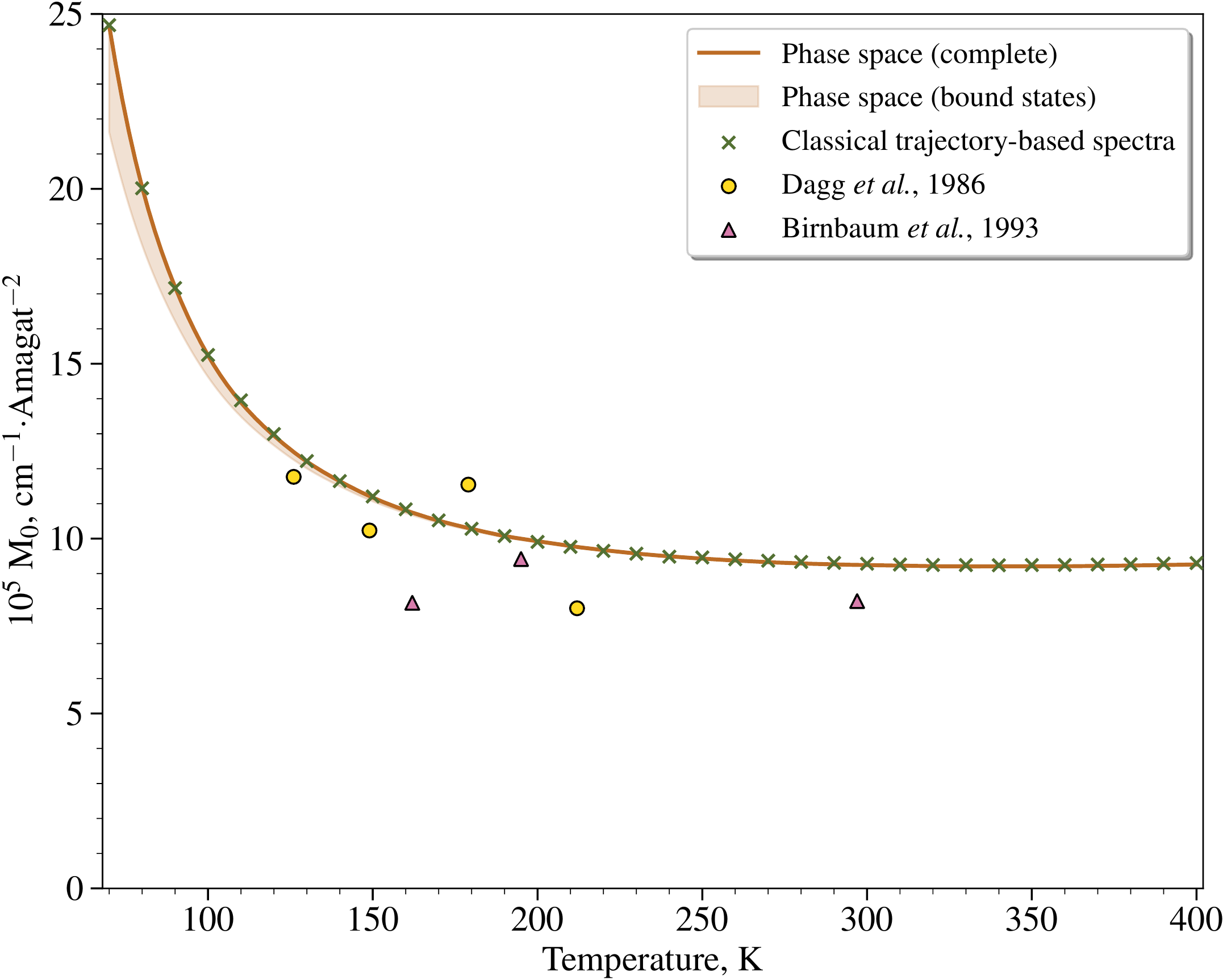}
    \includegraphics[width=0.49\linewidth]{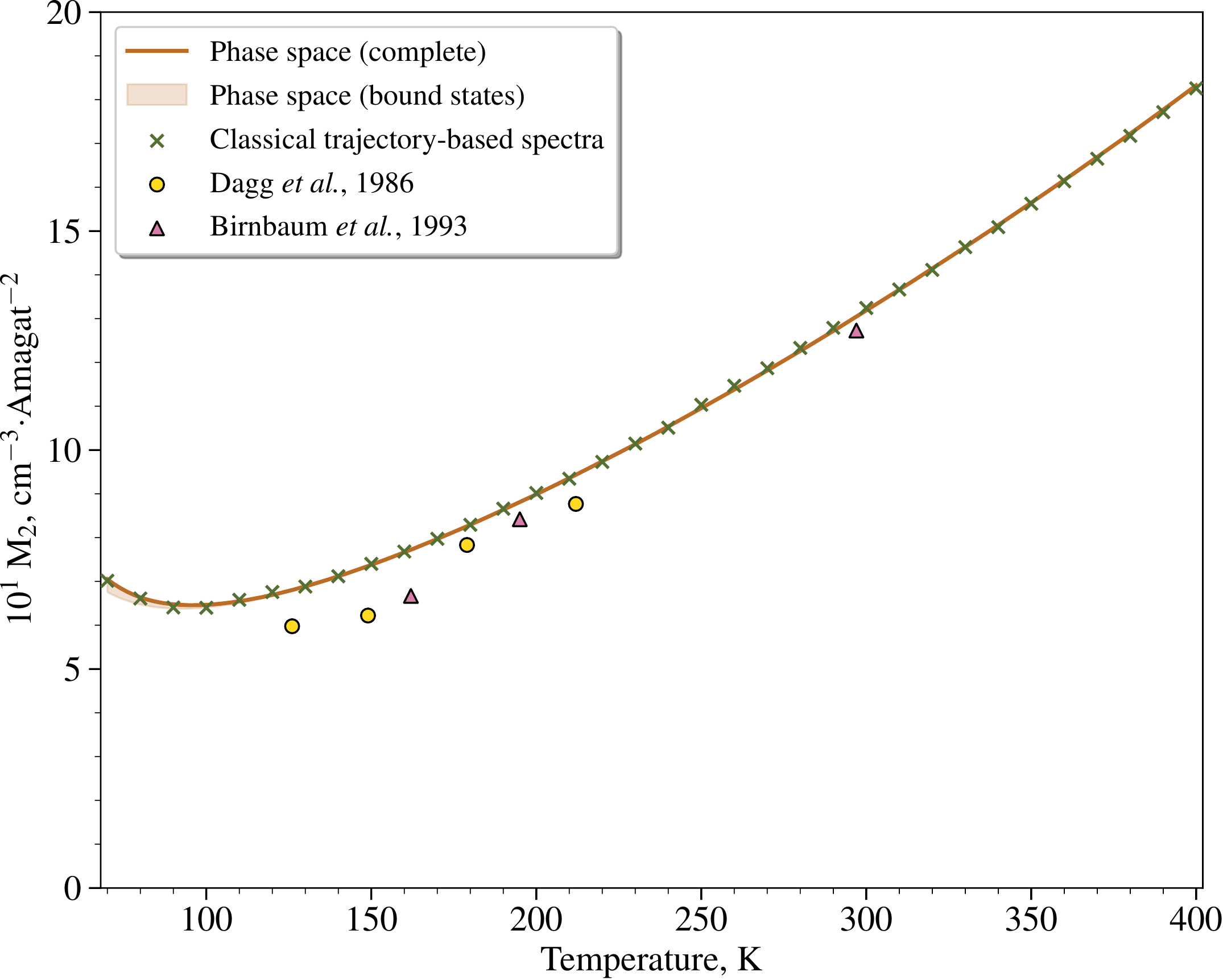}
    \caption{Calculated and experimental temperature variations of zeroth (left panel) and second (right panel) spectral moments of the CH$_4-$N$_2$ rototranslational band. The solid line refers to the values obtained through phase-space integration~\eqref{general:spectral-moments-phase-space}. The crosses show the values derived from the trajectory-based profiles using Eq.~\eqref{general:spectral-moments-alpha}. The shaded area indicates the true bound states' contribution. Circles and triangles represent the values retrieved from the measurements by \citet{Dagg1986} and \citet{Birnbaum1993}, respectively.}
    \label{fig:spectral-moments}
\end{figure*}

In Fig.~\ref{fig:spectral-moments}, classical spectral moments obtained through phase space integration~\eqref{general:spectral-moments-phase-space} are compared to the values derived from the trajectory-based spectral profiles via Eq.~\eqref{general:spectral-moments-alpha} as well as from the laboratory-measured spectra. Perfect agreement between the spectral moments obtained via frequency-domain and phase-space integration validates calculated spectral profiles.

\begin{figure*}
    \centering
    \includegraphics[width=0.49\linewidth]{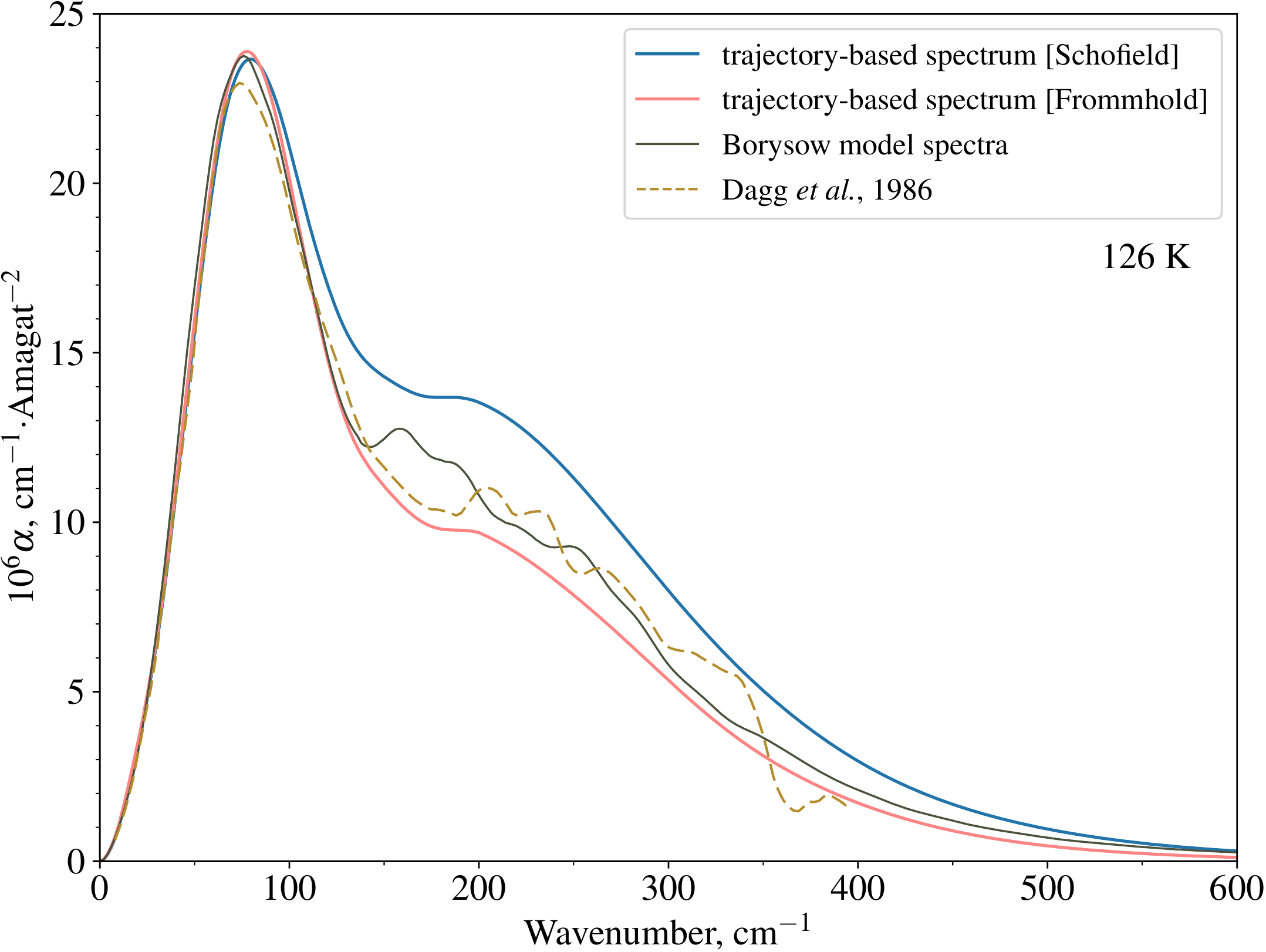}
    \includegraphics[width=0.49\linewidth]{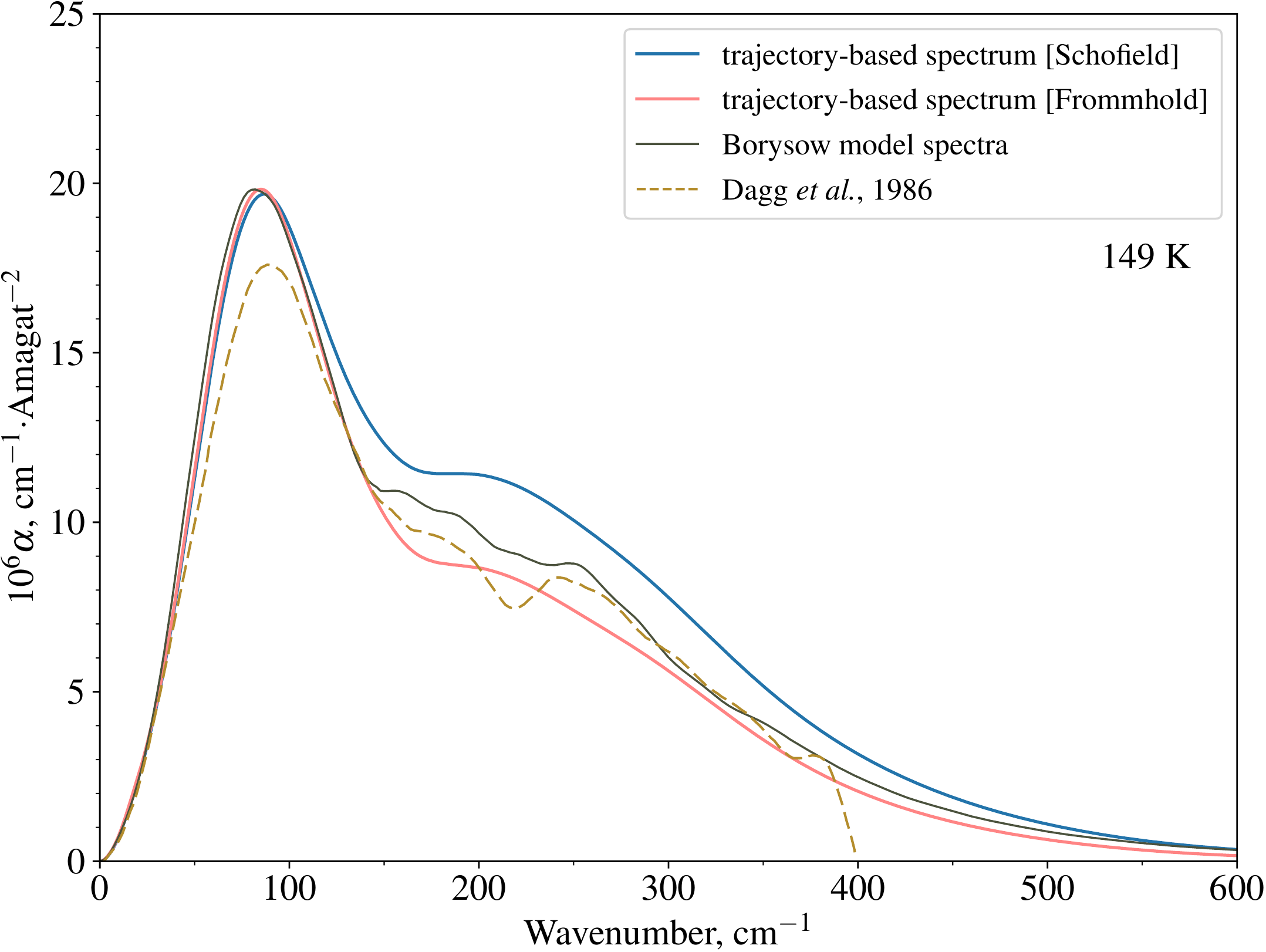} \\
    \includegraphics[width=0.49\linewidth]{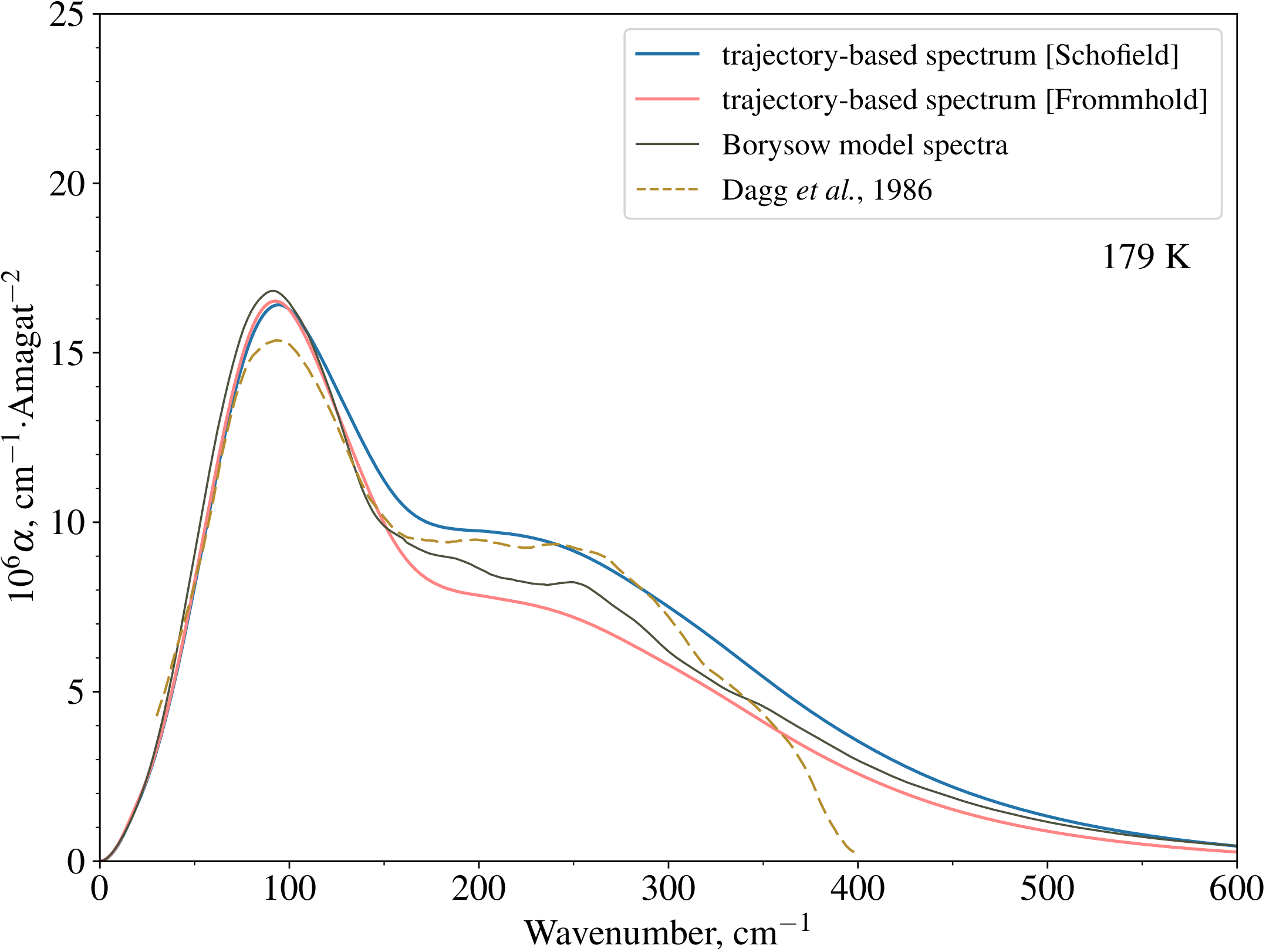}
    \includegraphics[width=0.49\linewidth]{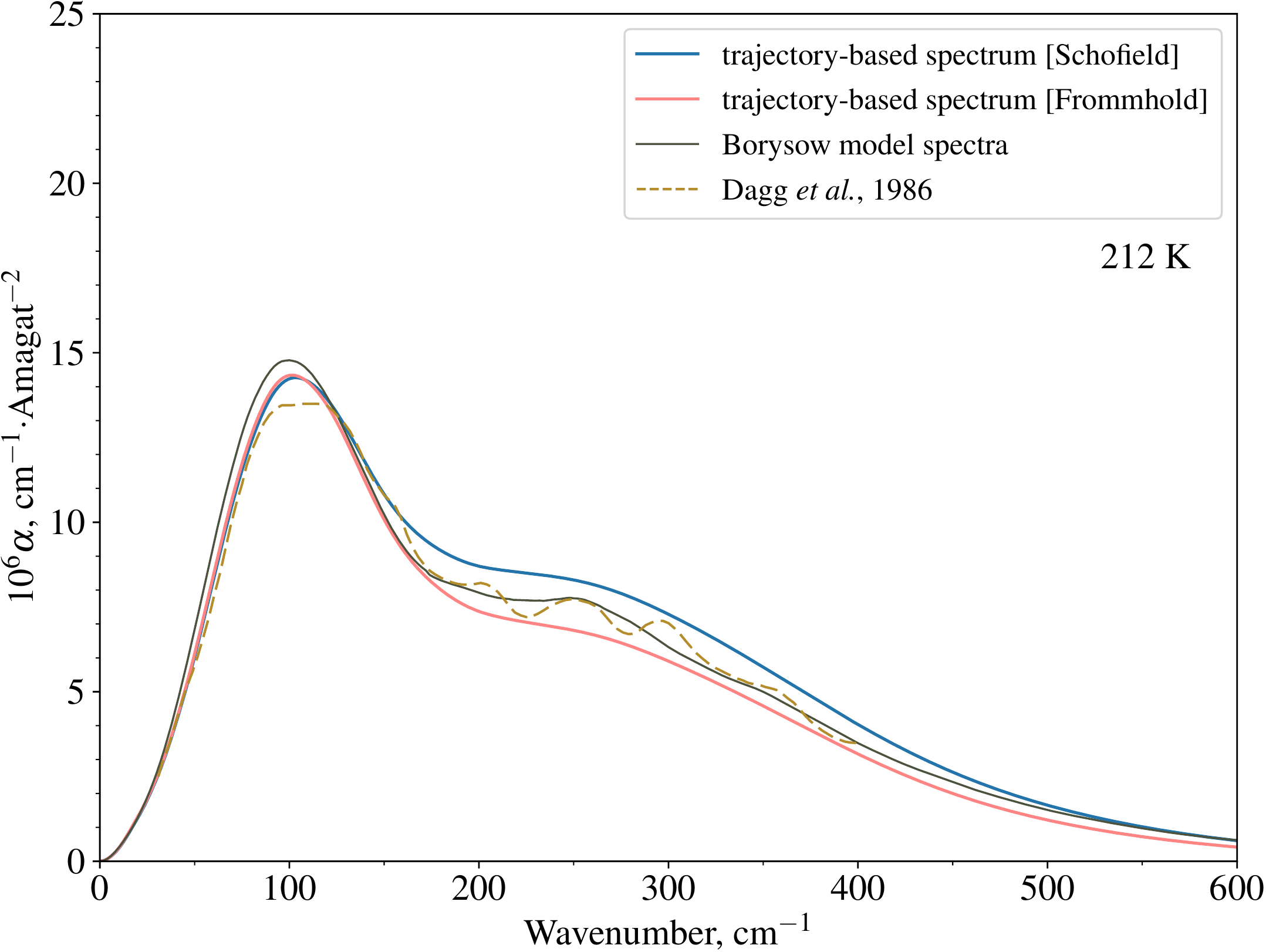}
    \caption{CH$_4-$N$_2$ collision-induced absorption spectra at selected temperatures. Dashed yellow lines denote experimental measurements by \citet{Dagg1986} at 126, 149, 179, and 212~K. The blue and red lines indicate the trajectory-based results obtained using Schofield's and Frommhold's quantum correction factors. Model profiles of \citet{Borysow1993} are represented with gray lines.}
    \label{fig:dagg-comparison-spectra}
\end{figure*}

\begin{figure*}
    \centering
    \includegraphics[width=0.49\linewidth]{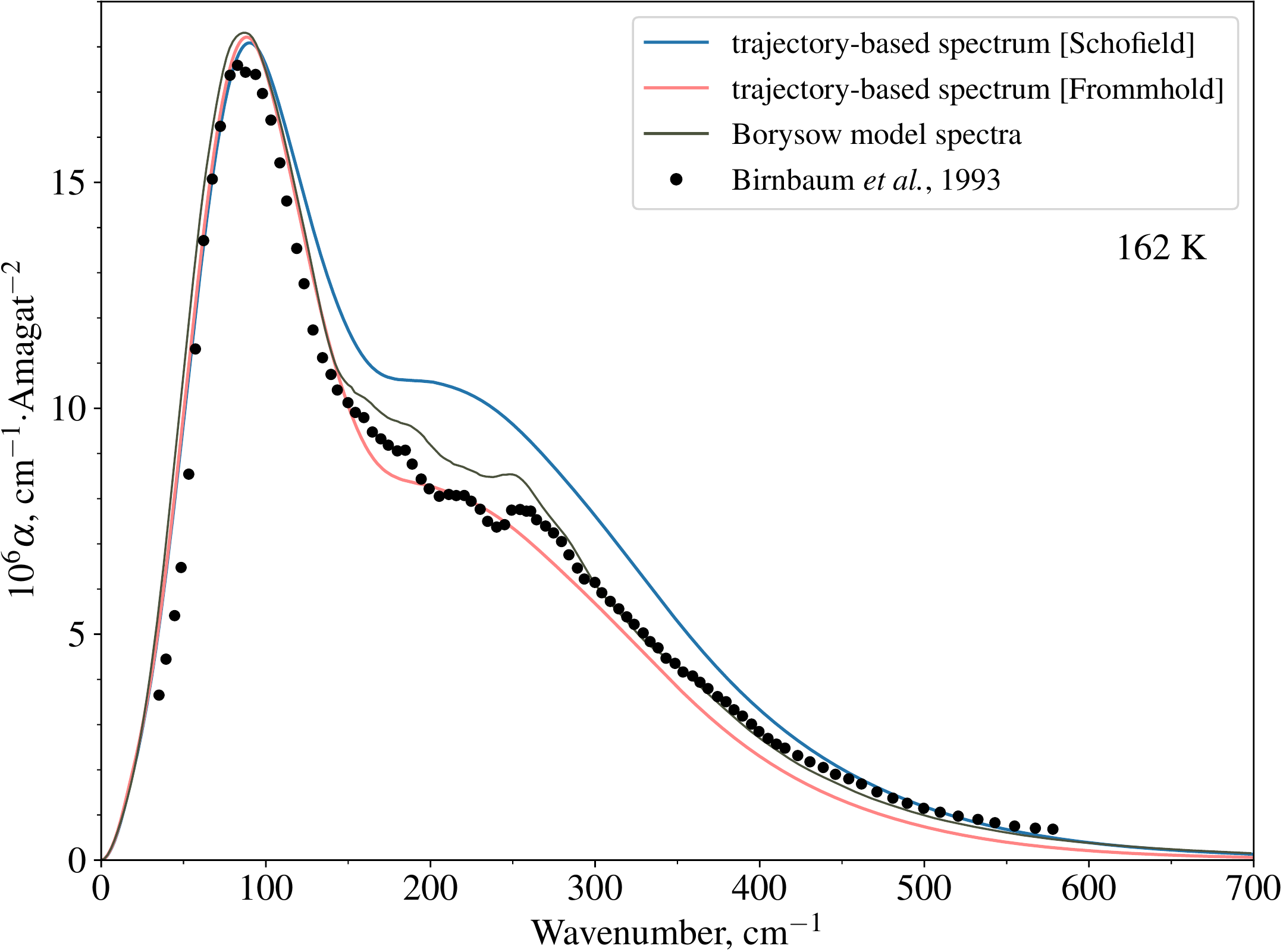}
    \includegraphics[width=0.49\linewidth]{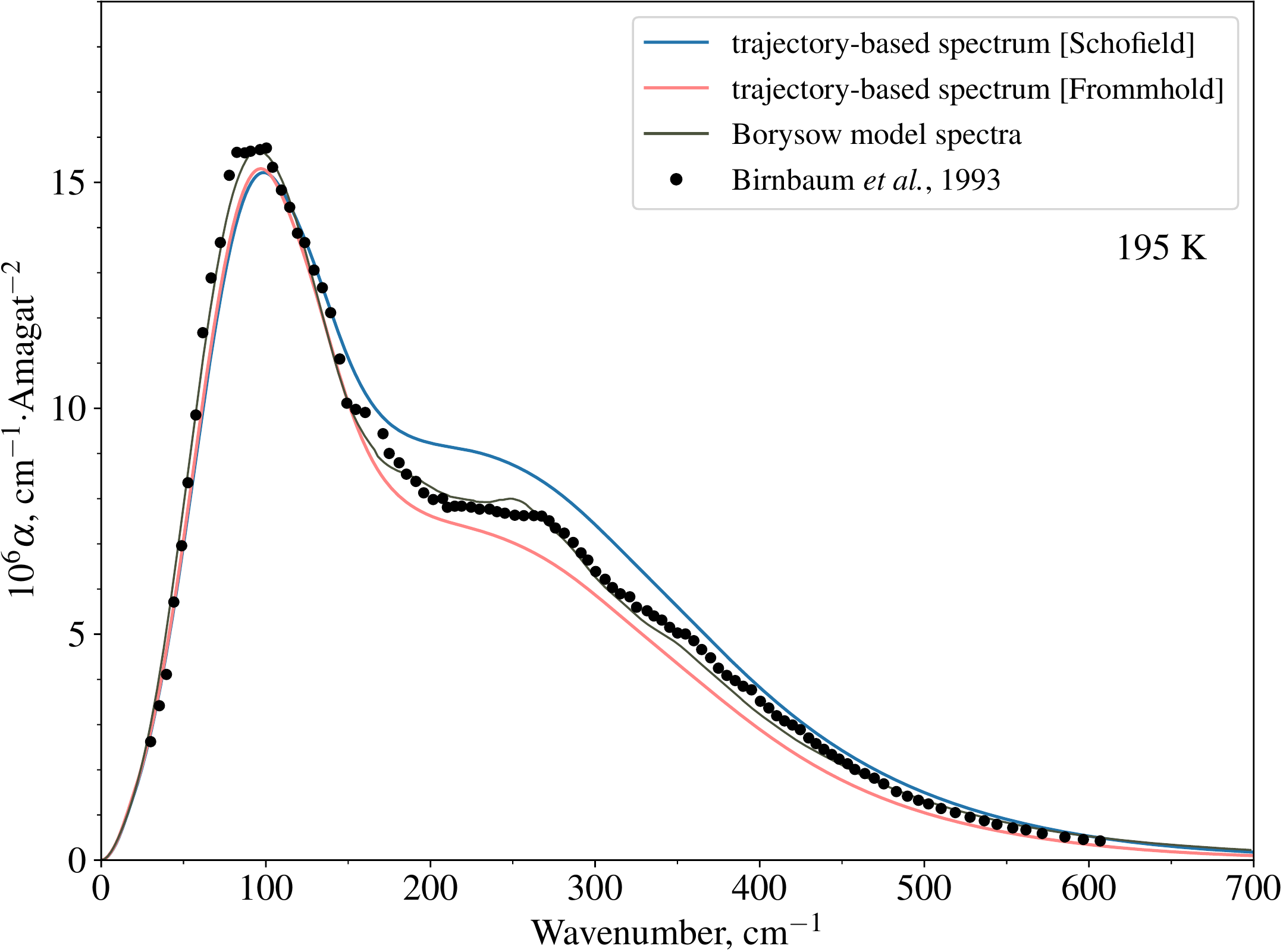} \\
    \includegraphics[width=0.49\linewidth]{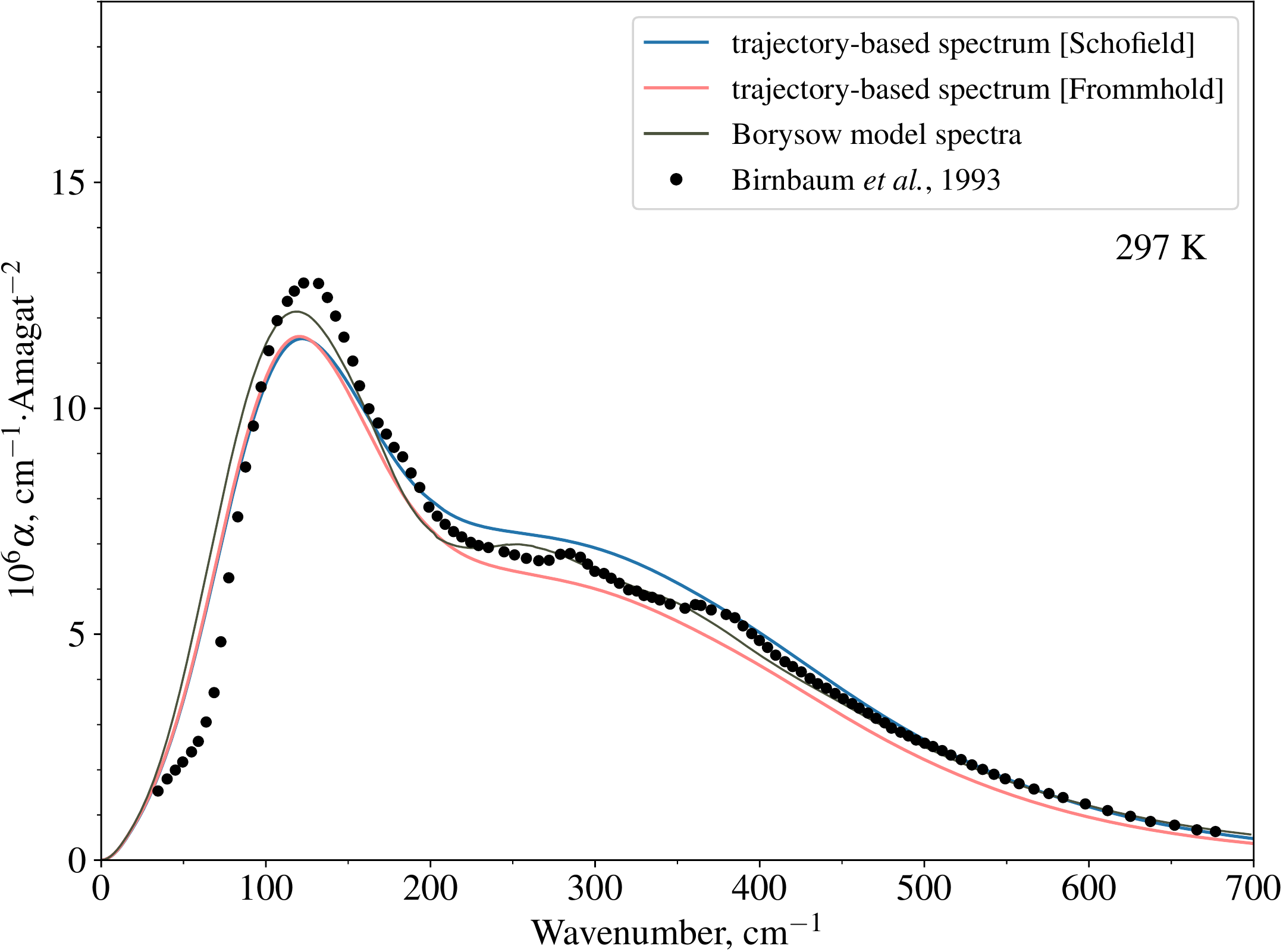}
    \caption{CH$_4-$N$_2$ collision-induced absorption spectra at selected temperatures. Markers denote experimental data by \citet{Birnbaum1993} at 162, 195, and 297~K. The blue and red lines indicate the trajectory-based results obtained using Schofield's and Frommhold's quantum correction factors. Model profiles of \citet{Borysow1993} are represented with gray lines.}
    \label{fig:birnbaum-comparison-spectra}
\end{figure*}

Figures~\ref{fig:dagg-comparison-spectra} and \ref{fig:birnbaum-comparison-spectra} compare the calculated CIA CH$_4-$N$_2$ spectral profiles with the experimental measurements by \citet{Dagg1986} and \citet{Birnbaum1993}, respectively. It can be seen that in the spectral shoulder and far wing, between 150 and 600 cm$^{-1}$, the spectral profiles desymmetrized using Schofield's and Frommhold's corrections act as upper and lower bounds on the experimental spectrum. However, the deviations between profiles corrected using Frommhold's procedure and experimental data are slightly less compared to those obtained using Schofield's procedure. 

%%%%%%%%%%%%%%%%%%%%%%%%%%%%%%%%%%%%%%%%%%%%%%%%%%%%%%%%%%%%%%%%%%%%%%%%%%%%%%%%%%%%%%%%%%%%%%%
\section{Modeling CIRS/Cassini measurements}
\label{sec:modeling-cassini-measurements}
%%%%%%%%%%%%%%%%%%%%%%%%%%%%%%%%%%%%%%%%%%%%%%%%%%%%%%%%%%%%%%%%%%%%%%%%%%%%%%%%%%%%%%%%%%%%%%%

%%%%%%%%%%%%%%%%%%%%%%%%%%%%%%%%%%%%%%%%%%%%%%%%%%%%%%%%%%%%%%%%%%%%%%%%%%%%%%%%%%%%%%%%%%%%%%%
\subsection{Observations}
\label{subsec:observations}
%%%%%%%%%%%%%%%%%%%%%%%%%%%%%%%%%%%%%%%%%%%%%%%%%%%%%%%%%%%%%%%%%%%%%%%%%%%%%%%%%%%%%%%%%%%%%%%

The Cassini/CIRS instrument consists of two interferometers, sharing a common telescope and scan mechanism. The far-infrared interferometer using Focal Plane 1 (hereafter referred to as FP1) covers the spectral range 10--650~cm$^{-1}$ with an adjustable apodized spectral resolution from 15.5 to as high as 0.5~cm$^{-1}$. This focal plane consists of two thermopile detectors with a 3.9-mrad circular field of view with a half-peak diameter of 2.5 mrad \citep{Flasar2004, Jennings2017}. Bearing in mind the time (14 January 2005) and location (10.3$^\circ$S, 167.7$^\circ$E) of the Huygens landing site, we focused on spectra recorded between 2005 and 2007 in the 20$^\circ$S--20$^\circ$N latitude range. Seasonal variations of temperature and gas abundances are weak in the considered latitude range \citep{Bezard2018}, allowing us to use the constraints derived from the \textit{in situ} measurements conducted by the Huygens probe. We opted to use the same two selections of 15.5~cm$^{-1}$ resolution spectra as those made by \citet{Bezard2020} to retrieve the H$_2$ mole fraction and ortho-to-para ratio. The two selections have different mean emission angles, ``low'' (13$^\circ$) and ``high'' (59$^\circ$), allowing us to more effectively disentangle the contributions from tropospheric CIA and stratospheric haze emission \citep{Samuelson1997}.

%%%%%%%%%%%%%%%%%%%%%%%%%%%%%%%%%%%%%%%%%%%%%%%%%%%%%%%%%%%%%%%%%%%%%%%%%%%%%%%%%%%%%%%%%%%%%%%
\subsection{Radiative transfer model}
\label{subsec:radiative-transfer-model}
%%%%%%%%%%%%%%%%%%%%%%%%%%%%%%%%%%%%%%%%%%%%%%%%%%%%%%%%%%%%%%%%%%%%%%%%%%%%%%%%%%%%%%%%%%%%%%%

Synthetic spectra were generated using the radiative transfer code described in \citet{Bezard2020} and including CIA from N$_2-$N$_2$, CH$_4-$N$_2$, H$_2-$N$_2$ and CH$_4-$CH$_4$, rotational lines of CO, HCN and CH$_4$, rovibrational bands from C$_4$H$_2$ and CH$_3$C$_2$H, and haze particles. The atmospheric model is the same as that of \citet{Bezard2020} and makes use of \textit{in situ} Huygens measurements of temperature \citep{Fulchignoni2005}, tropospheric methane abundance \citep{Niemann2010}, and main haze profile \citep{Doose2016}. Additional constraints on the aerosol opacity profile (including the nitrile ice cloud described by \citet{Anderson2011}) as well as C$_4$H$_2$ and CH$_3$C$_2$H abundance profiles were obtained from CIRS limb spectra (at viewing altitudes of 105 and 145 km) recorded in the same latitude and time ranges as the two surface-intercepting spectral selections \citep{Bezard2020}.

\begin{figure}
    \centering
    \includegraphics[width=\linewidth]{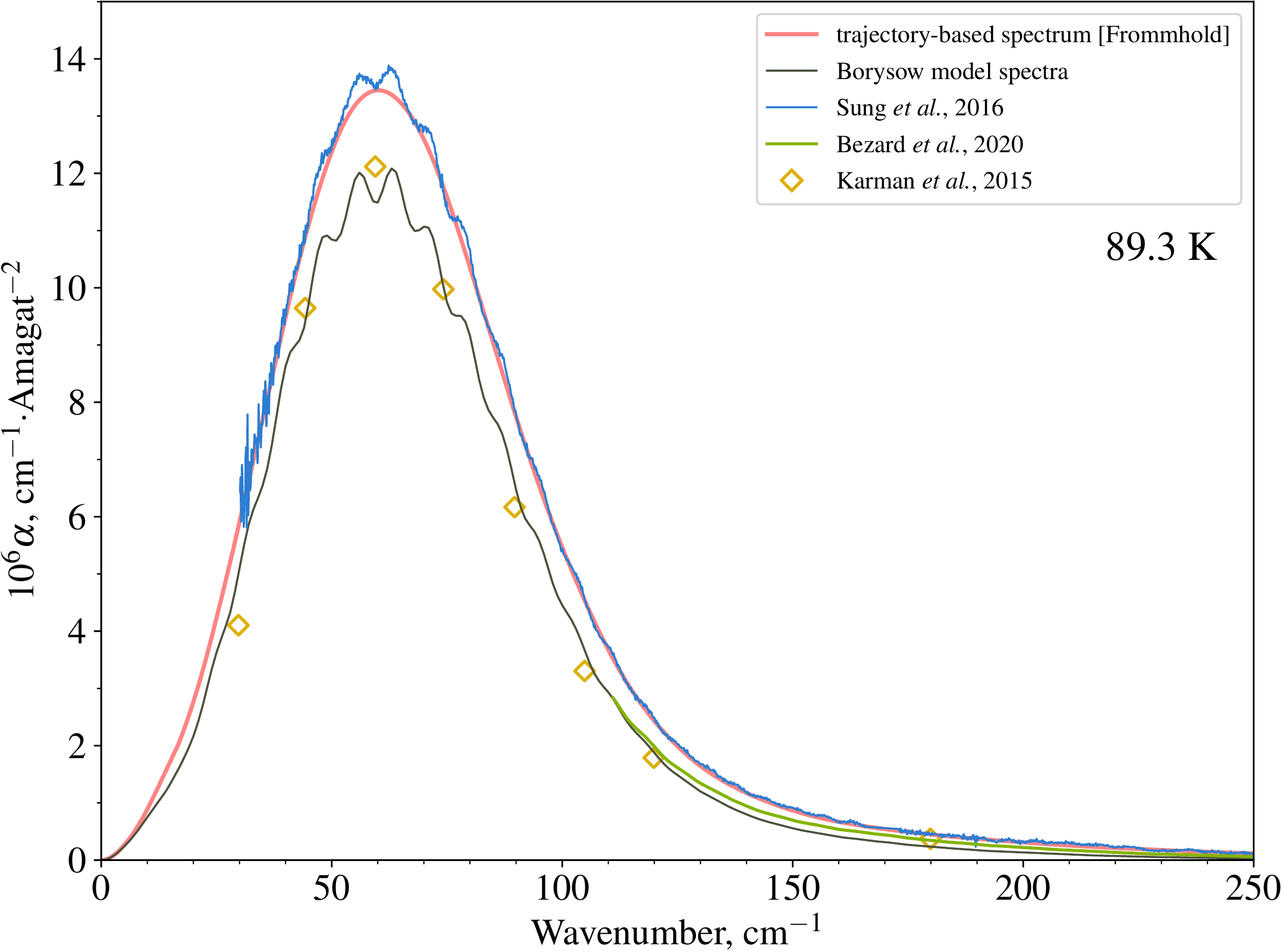}
    \caption{N$_2-$N$_2$ collision-induced absorption spectra at 89.3~K. The red curve indicates the results of the trajectory-based calculation (\citet{Chistikov2019}) utilizing Frommhold's desymmetrization procedure. The experimental measurements of \citet{Sung2016} are denoted with the blue line. The yellow diamonds and gray curve denote the results of quantum mechanical calculations of \citet{Karman2015} and \citet{Borysow1986-N2-N2}, respectively. The adjusted profile utilized in \citet{Bezard2020} is indicated with the green line.}
    \label{fig:n2-n2-89K}
\end{figure}

The N$_2-$N$_2$ and CH$_4-$N$_2$ CIA coefficients were updated with respect to \citet{Bezard2020}'s analysis. We opted for the use of the most recent N$_2-$N$_2$ CIA coefficients calculated using a trajectory-based approach \citep{Chistikov2019, HITRAN2020} utilizing \textit{ab initio} PES and IDS of \citet{Karman2015}. In contrast, \citet{Bezard2020} modelled the N$_2-$N$_2$ absorption by slightly adjusting the spectral profiles of \citet{Borysow1986-N2-N2}. They introduced a  wavenumber- and temperature-dependent multiplicative factor such that adjusted CIA coefficients in the far wing match the quantum mechanical calculations of \citet{Karman2015}.  Figure~\ref{fig:n2-n2-89K} compares the experimental measurements by \citet{Sung2016} at 89.3~K against the theoretical results by \citet{Chistikov2019}, \citet{Karman2015}, \citet{Borysow1986-N2-N2}, and heuristic model of \citet{Bezard2020}. It is seen that in the vicinity of the absorption peak, trajectory-based results of \citet{Chistikov2019} demonstrate significant improvement compared to previous theoretical efforts. In the far wing, neither of the calculations demonstrate perfect agreement with the experimental measurements of \citet{Sung2016} at temperatures 78.3, 89.3, 109.6, and 129.0~K. However, for 89.3 and 129.0~K, the deviations between the trajectory-based results and laboratory-measured spectra do not exceed 15\% over the rototranslational band.   

\begin{figure}
    \centering
    \includegraphics[width=\linewidth]{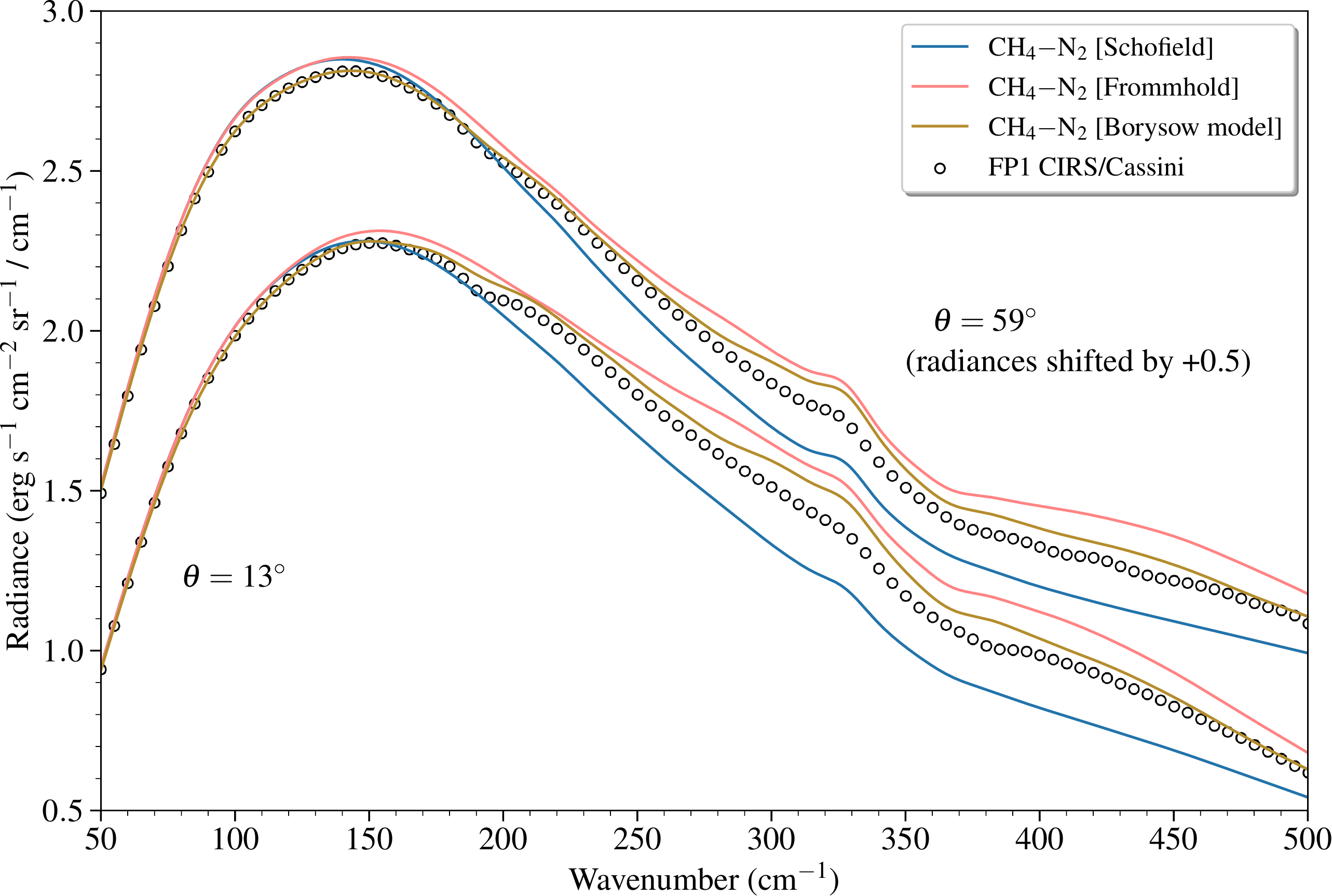}
    \caption{Far-infrared low (13$^\circ$) and high (59$^\circ$) emission angle spectra. CIRS FP1 radiances are represented with black circles. \myedit{The synthetic profiles using CH$_4-$N$_2$ CIA of \citet{Borysow1993} and trajectory-based results desymmetrized with Schofield's and Frommhold's procedures are denoted with yellow and blue and red lines, respectively.} The high-emission spectra are shifted by 0.5~erg\,s$^{-1}$\,cm$^{-2}$\,sr$^{-1}$/cm$^{-1}$ for clarity. \myedit{The minor emission feature near 330~cm$^{-1}$ is due to methyl-acetylene (CH$_3$C$_2$H), while the broad absorption centered at 355~cm$^{-1}$ is due to the S$_0(0)$ H$_2$ line.}}
    \label{fig:radiance-D3-D4a}
\end{figure}

\begin{figure}
    \centering
    \includegraphics[width=\linewidth]{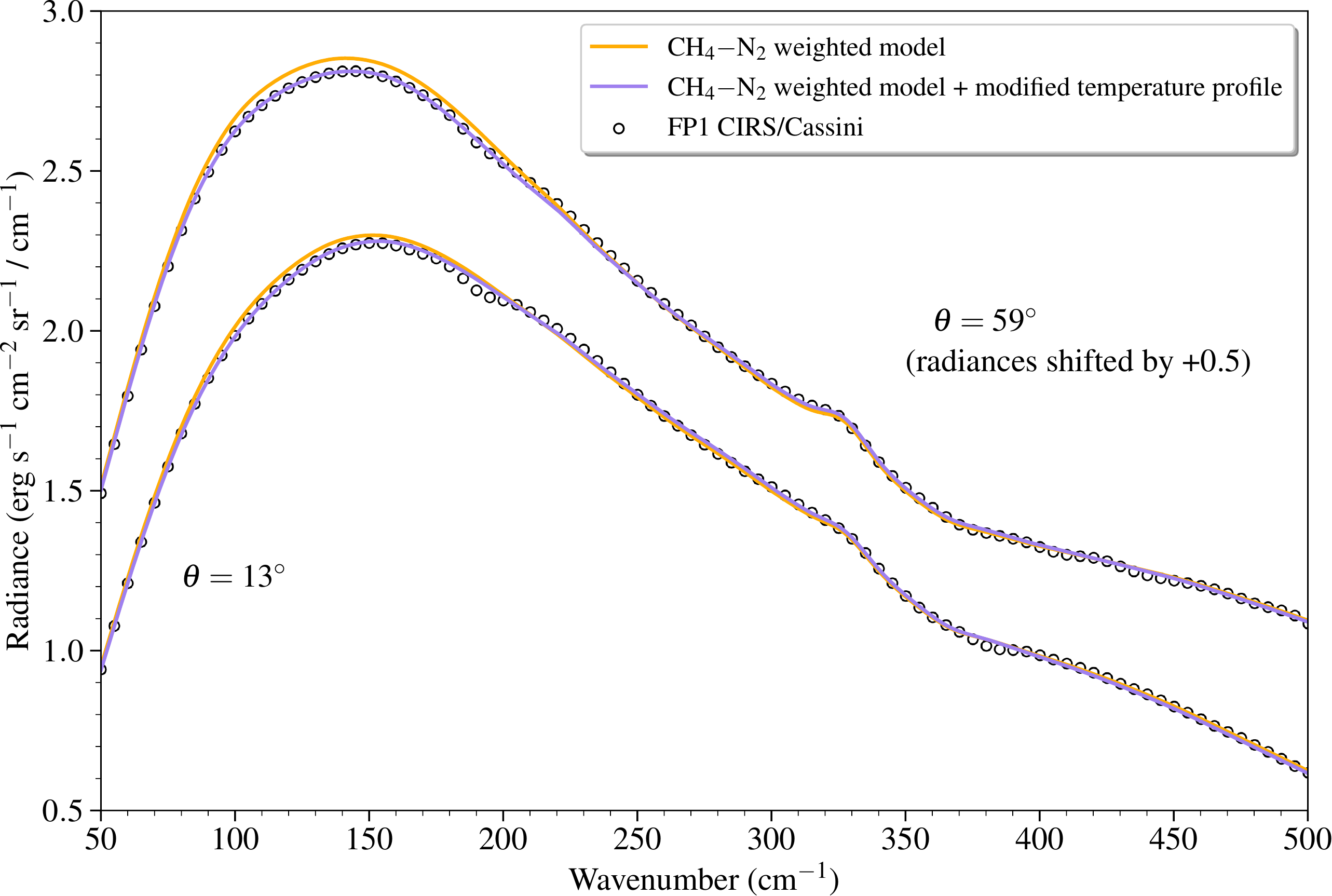}
    \caption{Far-infrared low (13$^\circ$) and high (59$^\circ$) emission angle spectra. CIRS FP1 radiances are represented with black circles. The synthetic profiles simulated using semi-empirical CH$_4-$N$_2$ CIA model~\eqref{modeling:linear-model} are denoted with orange lines. The radiances simulated using semi-empirical CIA CH$_4-$N$_2$, as well as modified temperature profile, are represented with magenta lines. The high-emission spectra are shifted by 0.5~erg\,s$^{-1}$\,cm$^{-2}$\,sr$^{-1}$/cm$^{-1}$ for clarity.}
    \label{fig:radiance-D3-D4-mixed}
\end{figure}

The analysis of spectral selections at low (13$^\circ$) and high (59$^\circ$) emission angles was performed using CH$_4-$N$_2$ CIA coefficients obtained for both quantum correction factors. The simulated radiances are shown in Fig.~\ref{fig:radiance-D3-D4a}. Bearing in mind the comparison of the calculated CH$_4-$N$_2$ CIA coefficients with the experimental measurements, it seems reasonable to consider a semi-empirical model in the form of linearly-weighted desymmetrized profiles
\begin{gather}
    \alpha_\text{SE}(\nu, T) = \Big[ c_1 \mathcal{F}_\text{Sch} \lb \nu, T \rb + c_2 \mathcal{F}_\text{Fromm} \lb \nu, T \rb \Big] \alpha_\text{cl} \lb \nu, T \rb,
    \label{modeling:linear-model}
\end{gather}
where coefficients $c_1, c_2$ are wavenumber- and temperature-independent. The best-fitting values of $c_1$ and $c_2$, which minimize the residuals between synthetic spectra and both CIRS data, were found to be 0.35 and 0.65, respectively. The fitted weights are congruent with the laboratory measurements, in particular, of \citet{Birnbaum1993} gravitating to trajectory-based profiles desymmetrized using Frommhold's procedure. Figure~\ref{fig:radiance-D3-D4-mixed} shows the synthetic radiances obtained using semi-empirical model~\eqref{modeling:linear-model}.   

The slight mismatch between synthetic radiances and CIRS data in the 100--200 cm$^{-1}$ range most likely indicates that the temperature profile we used \citep{Fulchignoni2005} is slightly too warm in the lower stratosphere and/or tropopause region. To investigate this possibility, we retrieved a temperature profile by inverting the CIRS radiances of both spectra in the range 55--545 cm$^{-1}$, using the Huygens probe profile \citep{Fulchignoni2005} as the \textit{a priori} and a correlation length of 0.75 scale height in the inversion algorithm. The so-derived temperature profile is colder than our initial profile by about 1 K in the 50--150 mbar region. We consider that this difference is physically acceptable given that the Huygens measurements were made at a single location and time while the CIRS spectra correspond to an average over 3 years and 40$^\circ$ in latitude. The synthetic spectra simulated with our semi-empirical CH$_4-$N$_2$ CIA model and retrieved temperature profile, shown in Figure~\ref{fig:radiance-D3-D4-mixed}, excellently reproduce the CIRS spectra over the spectral range 50--500~cm$^{-1}$.

%%%%%%%%%%%%%%%%%%%%%%%%%%%%%%%%%%%%%%%%%%%%%%%%%%%%%%%%%%%%%%%%%%%%%%%%%%%%%%%%%%%%%%%%%%%%%%%
\section{Implications for Archean Earth}
\label{sec:archean-earth}
%%%%%%%%%%%%%%%%%%%%%%%%%%%%%%%%%%%%%%%%%%%%%%%%%%%%%%%%%%%%%%%%%%%%%%%%%%%%%%%%%%%%%%%%%%%%%%%

\begin{figure*}
    \centering
    \includegraphics[width=0.49\linewidth]{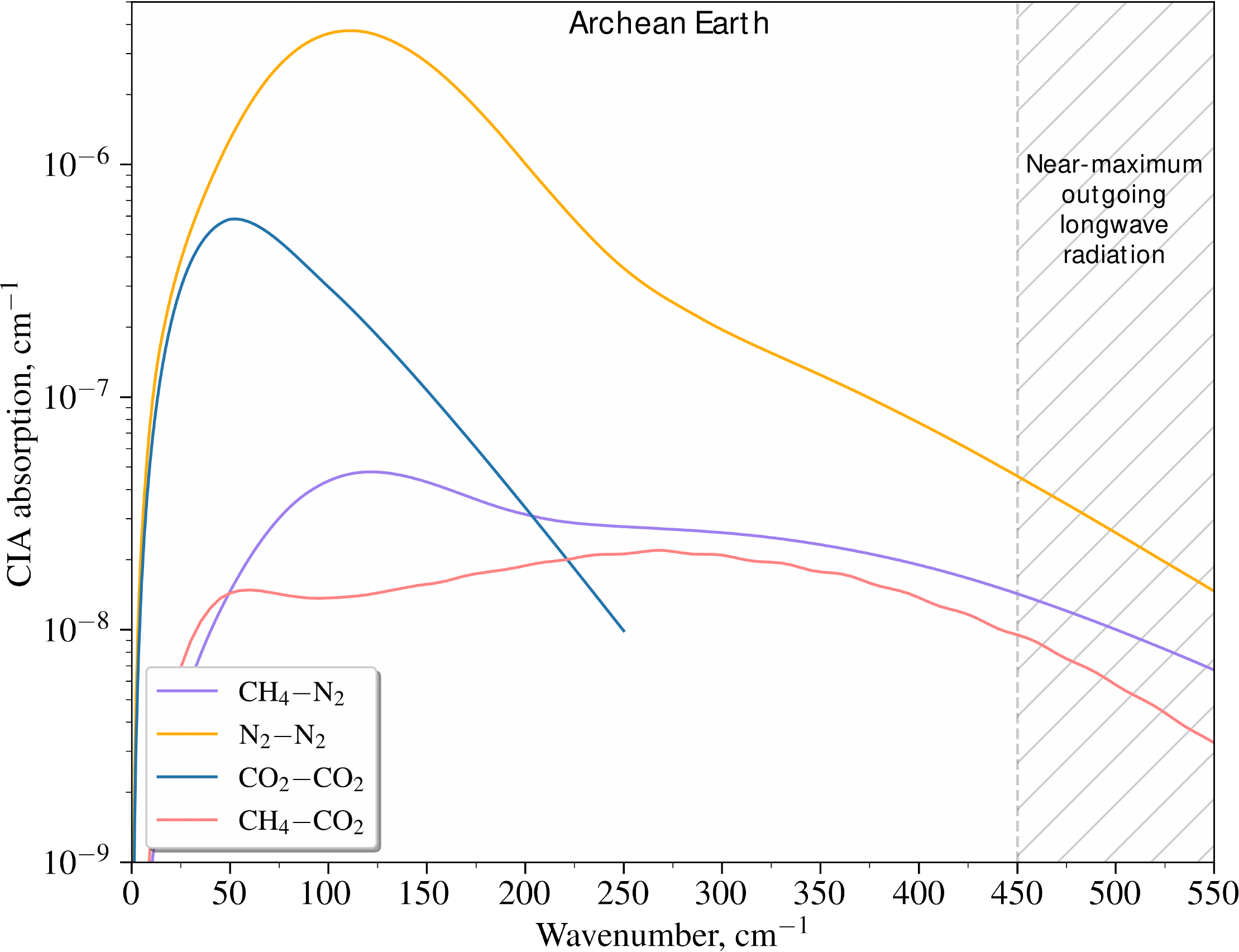}
    \includegraphics[width=0.49\linewidth]{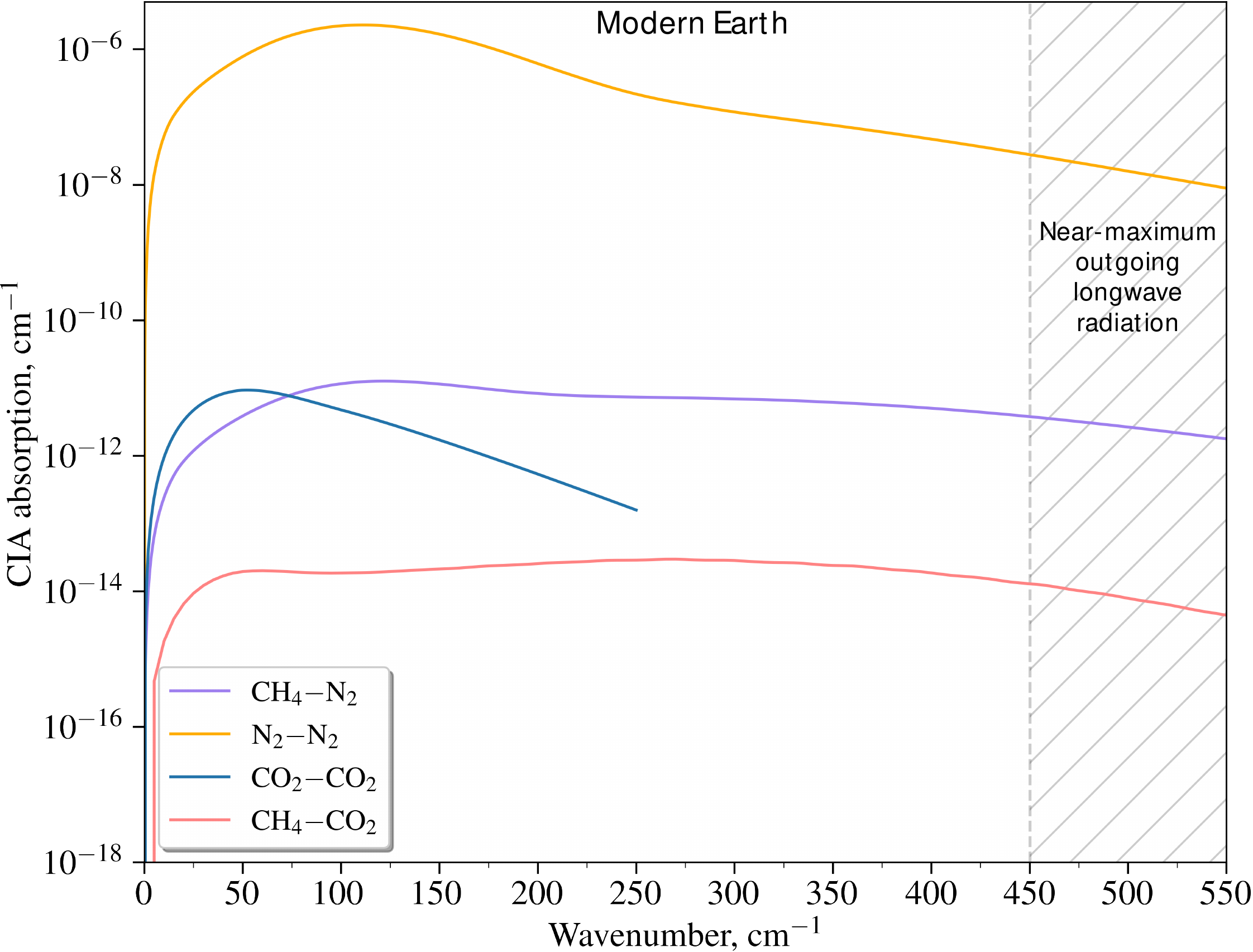}
    \caption{Far-infrared collision-induced absorption from selected pairs for the Archean atmosphere (left panel) with 1~atm of N$_2$, 0.1~atm of CO$_2$, $5 \cdot 10^{-3}$~atm of CH$_4$, and surface temperature of 300~K, and for the modern Earth conditions (right panel). The N$_2$ partial pressure is based on the constraint derived from the analysis of fluid inclusions in Archean rocks \citep{Avice2018}. The partial pressure of CO$_2$ is selected as the largest value predicted by the carbonate-silicate climate model of \citet{Krissansen-Totton2018-II}. Methane partial pressure is based on the constraint inferred from the fractionation of xenon isotopes \citep{Zahnle2019}.
    A mean global temperature of 300~K is assumed based on studies of the oxygen isotope ratios in sedimentary evidence \citep{Blake2010, Hren2009}. CIA coefficients for CO$_2-$CO$_2$, N$_2-$N$_2$, and CH$_4-$CO$_2$ are taken from \citet{Gruszka1997}, \citet{Chistikov2019}, and \citet{Wordsworth2017}, respectively. The semi-classical model (Eq.~\eqref{modeling:linear-model}) is utilized to calculate the CH$_4-$N$_2$ CIA contribution. Note significantly different scales on the panels.}
    \label{fig:archean}
\end{figure*}

Understanding the environmental conditions in the terrestrial atmosphere during the Archean eon -- from 4 to 2.5 billion years ago -- may shed light on the biological and geological evolution of our planet and Earth-like exoplanets \citep{Catling2020, Krissansen-Totton2018-I}. In the Archean, the atmosphere was devoid of oxygen, and the bulk gases, N$_2$ and CO$_2$, were supplemented with reducing gases such as CH$_4$ \citep{Catling2014}. High levels of biogenic methane, as well as hydrocarbon haze layers similar to Titan's, have been suggested to warm the early Earth \citep{Pavlov2001}. Pieces of evidence, such as pronounced fractionation of xenon isotopes \citep{Zahnle2019} and mass-independent fractionation of sulfur isotopes \citep{Zahnle2006}, point to high mixing ratios of methane, some reaching 0.5\%. In turn, it follows that interacting pairs of CH$_4-$N$_2$ and CH$_4-$CO$_2$ may need to be considered as potential contributors to greenhouse warming relevant in the context of the faint young Sun problem (see, \textit{e.g.}, a recent review by \citet{Charnay2020}). 
Collision-induced absorption is the major component of the opacity of gas giants and Titan in the dense regions of the atmosphere. However, its implications for early Earth warming are not well studied. While N$_2-$N$_2$ and CO$_2-$CO$_2$ absorption is strong shortward of 300~cm$^{-1}$, it is too far from the peak of black-body emission at terrestrial temperatures to have a significant greenhouse effect. The CH$_4-$N$_2$ and CH$_4-$CO$_2$ CIA bands are broader and may be important in the critical 450--700~cm$^{-1}$ range near the maximum of outgoing radiation, where the contribution of other gases to opacity is small. Figure~\ref{fig:archean} shows the N$_2-$N$_2$, CH$_4-$N$_2$, CO$_2-$CO$_2$, and CH$_4-$CO$_2$ collision-induced components assuming a model Archean and modern atmospheres. As illustrated, the opacities due to methane-containing pairs were significantly larger in the Archean atmosphere than in the modern one. Despite that, it can be concluded that the CIA due to CH$_4-$N$_2$ and CH$_4-$CO$_2$ pairs is of little importance for infrared radiative transfer in the Archean atmosphere taking the constraint for CH$_4$ partial pressure inferred from the fractionation of xenon isotopes \citep{Zahnle2019}. Our results corroborate the findings of \citet{Pavlov2000}, who concluded that CH$_4-$N$_2$ CIA coefficients parameterized by \citet{Borysow1993}'s model are insignificant for greenhouse warming of early Earth. Nevertheless, should the constraints for methane mixing ratio be revised, the contribution of methane-containing intermolecular pairs may need to be revisited.

%%%%%%%%%%%%%%%%%%%%%%%%%%%%%%%%%%%%%%%%%%%%%%%%%%%%%%%%%%%%%%%%%%%%%%%%%%%%%%%%%%%%%%%%%%%%%%%
\section{Conclusions}
\label{sec:conclusions}
%%%%%%%%%%%%%%%%%%%%%%%%%%%%%%%%%%%%%%%%%%%%%%%%%%%%%%%%%%%%%%%%%%%%%%%%%%%%%%%%%%%%%%%%%%%%%%%

This work presents an extensive trajectory-based examination of the CH$_4-$N$_2$ rototranslational collision-induced absorption band. Our first-principles analysis relies on the approach presented in \citet{Chistikov2021}, allowing us to simulate the spectral contributions corresponding to both bound and unbound pair states. Significant progress has been made with respect to previous consideration of CH$_4-$N$_2$ CIA by \citet{Borysow1993} in virtue of the use of high-level \textit{ab initio} anisotropic PES and IDS \citep{Finenko2021}, obviating the need for short-range empirical dipole terms. We took advantage of the possibility to represent the lower-order spectral moments as phase-space and frequency-domain integrals to utilize them as the convergence control parameters of our Monte Carlo approach.

Classically derived trajectory-based profiles were subject to Schofield's \eqref{general:D3-correction} and Frommhold's \eqref{general:D4a-correction} desymmetrization procedures. These profiles were then compared to available experimental data in two ways. First, we demonstrate an overall satisfactory agreement of profiles issued from both desymmetrization procedures with the spectra measured by \citet{Birnbaum1993} and \citet{Dagg1986}. Second, the simulated profiles were utilized for modeling Titan's spectra in the 50--500~cm$^{-1}$ range. A semi-empirical model in the form of linearly-weighted desymmetrized profiles was designed to reproduce Cassini CIRS spectra at low and high emission angles. The synthetic spectra simulated with semi-empirical CH$_4-$N$_2$ CIA spectra and a slightly modified temperature profile yielded an excellent agreement with CIRS data. For that reason, we believe that the semi-empirical CH$_4-$N$_2$ CIA spectra we obtained are the most accurate to date and may be beneficial for future Titan studies. The semi-empirical CH$_4-$N$_2$ CIA profiles from 70 to 400~K with a step of 10~K tabulated in the HITRAN CIA format (see \citet{Karman2019} and HITRAN CIA website\footnote{\url{https://hitran.org/cia/}}) are provided in the Supplementary material. These results are proposed to be included in the subsequent edition of the CIA section of the HITRAN database \citep{HITRAN2020}.  

Since existing laboratory measurements of CH$_4-$N$_2$ CIA are limited to temperatures above 126~K \citep{Dagg1986}, we encourage laboratory studies down to the lowest available temperature to validate our semi-empirical and simulated profiles. Naturally, since our calculations were carried out only up to 400~K and there are no experimental data above that temperature, higher temperature experiments would also be very welcome to account for possible scenarios in exoplanetary atmospheres. 

We explored the potential implications of the simulated CH$_4-$N$_2$ CIA profiles for greenhouse warming of early Earth's atmosphere. Based on the current constraints for Archean levels of methane mixing ratio, we concluded that CH$_4-$N$_2$ CIA coefficients are insignificant for far-infrared radiative transfer in the Archean atmosphere, corroborating the findings of \citet{Pavlov2000}.

The highly-reduced atmospheres on rocky exoplanets are theoretically supported, assuming reduced iron can be retained in the mantle \citep{Lichtenberg2021}. The atmospheres of many Super-Earth exoplanets may resemble a warm Titan composition. For instance, \citet{Turbet2018} performed climate simulations to analyze the possibility of TRAPPIST-1 outer planets \citep{Gillon2017} being warm Titans. We believe that our CH$_4-$N$_2$ CIA profiles could play an important role in radiative transfer models for the climate simulations for warm N$_2$/CH$_4$ exoplanetary atmospheres.  

Trajectory-based calculations were carried out using HPC resources of Smithsonian Institution High Performance Cluster (SI/HPC)\footnote{\url{https://doi.org/10.25572/SIHPC}}. This work is supported by the NASA grants 80NSSC20K0962 and 80NSSC20K1059.

%%%%%%%%%%%%%%%%%%%%%%%%%%%%%%%%%%%%%%%%%%%%%%%%%%%%%%%%%%%%%%%%%%%%%%%%%%%%%%%%%%%%%%%%%%%%%%%
\appendix
\section*{A. Transformation of angles}
\refstepcounter{appendixcounter}
\label{appendix:transformation-of-angles}
%%%%%%%%%%%%%%%%%%%%%%%%%%%%%%%%%%%%%%%%%%%%%%%%%%%%%%%%%%%%%%%%%%%%%%%%%%%%%%%%%%%%%%%%%%%%%%%

\def\theequation{A.\arabic{equation}}
\setcounter{equation}{0}
\noindent

The components of any vector $\mf{v}$ in the space-fixed system $\mf{v}_\textup{SF}$ are linearly related to the components of the same vector in the body-fixed reference frame $\mf{v}_\textup{BF}$ through orthogonal transformation matrix $\bbS$ \citep{Goldstein2001}
\begin{gather}
    \bbS \mf{v}_\textup{SF} = \mf{v}_\textup{BF}.
    \label{app:SF-BF}
\end{gather}
The elements of the matrix $\bbS$ can be expressed in terms of three Euler angles $\lb \varphie, \varthetae, \psie \rb$. In the present paper, we adhere to the definition of Euler angles given by \citet{Goldstein2001} leading to the following form of the transformation matrix $\bbS$
\begin{gather}
    \bbS(\varphie, \varthetae, \psie) = \bbSz \lb \psie \rb \bbSx \lb \varthetae \rb \bbSz \lb \varphie \rb,
    \label{app:s-matrix}
\end{gather}
where a matrix $\bbS_\xi(\zeta)$ implements a counterclockwise rotation on angle $\zeta$ around $\xi$-axis. Rotation matrices about the $X$-, $Y$- and $Z$-axes are defined as
\begin{gather}
    \bbSx(\zeta) = \begin{bmatrix}
        1 & 0 & 0 \\
        0 & \cos \zeta & \sin \zeta \\
        0 & -\sin \zeta & \cos \zeta
    \end{bmatrix}, \quad
    \bbSy(\zeta) = \begin{bmatrix}
        \cos \zeta & 0 & -\sin \zeta \\
        0 & 1 & 0 \\
        \sin \zeta & 0 & \cos \zeta
    \end{bmatrix}, \quad
    \bbSz(\zeta) = \begin{bmatrix}
        \cos \zeta & \sin \zeta & 0 \\
        -\sin \zeta & \cos \zeta & 0 \\
        0 & 0 & 1
    \end{bmatrix}.
    \label{app:s-matrices}
\end{gather}

The spatial configuration of the CH$_4-$N$_2$ complex can be given in either body-fixed or space-fixed coordinate systems. To propagate the trajectory, one should be able to express body-fixed angles in terms of the space-fixed ones for the same configuration of the moieties since PES and IDS are expressed in terms of the body-fixed coordinates.
The basic idea of the procedure is that given the initial reference orientation of the moieties, an arbitrary configuration can be reached by the sequence of rotations through either body-fixed or spaced-fixed angles expressed as the product of rotation matrices. We select two vectors embedded in the complex configuration, and by equating their coordinates in body-fixed and space-fixed frames, we can derive the values of the body-fixed coordinates. For the intermolecular vector $\mathbf{R}$ we have  
\begin{gather}
    \mathbf{R}_\textup{SF} = R 
    \begin{bmatrix}
        \cos \Phi \sin \Theta \\
        \sin \Phi \sin \Theta \\
        \cos \Theta
    \end{bmatrix},
\end{gather}
and
\begin{gather}
    \mathbf{R}_\textup{BF} = R 
    \begin{bmatrix} 
        \cos \Phi^\text{BF} \sin \Theta^\text{BF} \\
        \sin \Phi^\text{BF} \sin \Theta^\text{BF} \\
        \cos \Theta^\text{BF}
    \end{bmatrix}
    .
\end{gather}
Bearing in mind Eq.~\eqref{app:SF-BF} and rewriting the spherical coordinates in terms of the rotation matrices, we obtain
\begin{gather}
    \bbSzt(\varphie) \bbSxt(\varthetae) \bbSzt(\psie) \bbSzt(\Phi^\text{BF}) \bbSyt(\Theta^\text{BF}) \unitvector = \bbSzt(\Phi) \bbSyt(\Theta) \unitvector.
\end{gather}
Note that the Euler angles $\lb \varphi, \vartheta, \psi \rb$ defining the orientation of the body-fixed frame with respect to the space-fixed frame are the same as the ones that describe the orientation of the CH$_4$ molecule in the space-fixed frame. Defining the auxiliary vector
\begin{gather}
    \mf{e}_1 = \bbSz(\psie) \bbSx(\varthetae) \bbSz(\varphie) \bbSzt(\Phi), \bbSyt(\Theta) \unitvector,
\end{gather}
one can obtain the body-fixed angles $\Theta^\text{BF}$ and $\Phi^\text{BF}$ through the following relations 
\begin{gather}
    \Theta^\text{BF}= \arccos \lb \mf{e}_1 \rb_Z, \quad 
    \Phi^\text{BF} = \atantwo \lb \frac{\lb \mf{e}_1 \rb_Y}{\sin \Theta^\text{BF}}, \frac{\lb \mf{e}_1 \rb_X}{\sin \Theta^\text{BF}} \rb,
\end{gather}
where $\atantwo$ is the 2-argument arctangent, which takes into account the signs of both arguments to determine the quadrant of the result.

Similarly, let us consider the vector $\mf{l}$ directed along the N$_2$ molecule. Its coordinates in the space-fixed and body-fixed frames of references are given by
\begin{gather}
    \mf{l}_\text{SF} = l
    \begin{bmatrix}
        \cos \eta \sin \chi \\
        \sin \eta \sin \chi \\
        \cos \chi
    \end{bmatrix}
\end{gather}
and 
\begin{gather}
    \mf{l}_\text{BF} = l 
    \begin{bmatrix} 
        \cos \eta^\text{BF} \sin \chi^\text{BF} \\
        \sin \eta^\text{BF} \sin \chi^\text{BF} \\
        \cos \chi^\text{BF}
    \end{bmatrix},
\end{gather}
respectively, where $l$ is the length of the N$_2$ molecule. Keeping in mind \eqref{app:SF-BF}, we have

\begin{gather}
    \bbSzt(\varphie) \bbSxt(\varthetae) \bbSzt(\psie) \bbSzt(\eta^\text{BF}) \bbSyt(\chi^\text{BF}) \unitvector = \bbSzt(\eta) \bbSyt(\chi) \unitvector.
\end{gather}
The body-fixed angles $\chi^\text{BF}$ and $\eta^\text{BF}$ angles can be obtained according to
\begin{gather}
    \chi^\text{BF} = \arccos \lb \mf{e}_2 \rb_Z, \quad 
    \eta^\text{BF} = \atantwo \lb \frac{\lb \mf{e}_2 \rb_Y}{\sin \chi^\text{BF}}, \frac{\lb \mf{e}_2 \rb_X}{\sin \chi^\text{BF}} \rb,
\end{gather}
where the auxiliary vector $\mf{e}_2$ is given by 
\begin{gather}
    \mf{e}_2 = \bbSz(\psie) \bbSx(\varthetae) \bbSz(\varphie) \bbSzt(\eta) \bbSyt(\chi) \unitvector.
\end{gather}

%Using the suggested scheme one can also easily calculate the Jacobian matrix of the derivatives of body-fixed angles over the spaced-fixed ones which is also necessary for trajectory propagation.
To compute the derivatives of the PES with respect to the space-fixed angles entering dynamical equations~\eqref{general:dynamical-equations}, we opt to employ the chain rule
\begin{gather}
    \frac{\partial U}{\partial \xi} = \sum_{\zeta =  \Phi^\text{BF}, \Theta^\text{BF}, \eta^\text{BF}, \chi^\text{BF}} \frac{\partial U}{\partial \zeta} \frac{\partial \zeta}{\partial \xi}, \quad \xi = \Phi, \Theta, \varphi, \vartheta, \psi, \eta, \chi.
\end{gather}
The required derivatives of the space-fixed angles with respect to the body-fixed can be readily obtained utilizing the same approach described above by addressing matrix calculus. Let us introduce additional auxiliary vectors
\begin{gather}
    \mf{e}_1^\zeta = \frac{\p}{\p \zeta} \Bigg[ \bbSz(\psie) \bbSx(\varthetae) \bbSz(\varphie) \bbSzt(\Phi) \bbSyt(\Theta) \Bigg] \unitvector, \\
    \mf{e}_2^\zeta = \frac{\p}{\p \zeta} \Bigg[ \bbSz(\psie) \bbSx(\varthetae) \bbSz(\varphie) \bbSzt(\eta) \bbSyt(\chi)  \Bigg] \unitvector, 
\end{gather}
where $\zeta = \{\Phi, \Theta, \varphi, \vartheta, \psi, \eta, \chi\}$ and $\partial / \partial \zeta$ means a standard matrix derivative \citep{Lax2007}. Then the corresponding derivatives can be computed via
\begin{gather}
    \frac{\p \Theta^\text{BF}}{\p \zeta} = - \frac{1}{\sin \Theta^\text{BF}} \lb \mf{e}_1^d \rb_Z, \quad 
    \frac{\p \Phi^\text{BF}}{\p \zeta} = \frac{1}{\sin \Theta^\text{BF}} \Bigg[ \lb \mf{e}_1^d \rb_Y \cos \Phi^\text{BF} - \lb \mf{e}_1^d \rb_X \sin \Phi^\text{BF} \Bigg], \\
   \frac{\p \chi^\text{BF}}{\p \zeta} = -\frac{1}{\sin \chi^\text{BF}} \lb \mf{e}_2^d \rb_Z, \quad 
   \frac{\p \eta^\text{BF}}{\p \zeta} = \frac{1}{\sin \chi^\text{BF}} \Bigg[ \lb \mf{e}_2^d \rb_Y \cos \eta^\text{BF} - \lb \mf{e}_2^d \rb_X \sin \eta^\text{BF} \Bigg].
  \end{gather}

%%%%%%%%%%%%%%%%%%%%%%%%%%%%%%%%%%%%%%%%%%%%%%%%%%%%%%%%%%%%%%%%%%%%%%%%%%%%%%%%%%%%%%%%%%%%%%%
\appendix
\section*{B. Initial condition generation}
\refstepcounter{appendixcounter}
\label{appendix:initial-condition-generation}
%%%%%%%%%%%%%%%%%%%%%%%%%%%%%%%%%%%%%%%%%%%%%%%%%%%%%%%%%%%%%%%%%%%%%%%%%%%%%%%%%%%%%%%%%%%%%%%

\def\theequation{B.\arabic{equation}}
\setcounter{equation}{0}
\noindent

In what follows, we outline the sampling algorithm from the Boltzmann distribution derived under the assumption of zero intermolecular potential
\begin{gather}
    \rho_0 \lb \mf{q}, \mf{p} \rb \sim \exp \lb - \beta \lb K_\text{tr} + K_{\text{rot}_1} + K_{\text{rot}_2} \rb \rb.
\end{gather}

Let us recall that the kinetic energy function is a quadratic form and can therefore be diagonalized (see, \textit{e.\,g.}, \citet{Chistikov2018})
\begin{gather}
    K_\text{tr} + K_{\text{rot}_1} + K_{\text{rot}_2} = \frac{1}{2} \mf{x}^\top \mf{x} = \frac{1}{2} \sum_{k = 1}^{8} x_k^2,
\end{gather}
where the relationship between generalized momenta $\mf{p}$ and variables $\mf{x}$ is given by
\begin{gather}
    \mf{p} = \bbT^\top \mf{x} = \begin{bmatrix}
        \sqrt{\mu} & 0 & 0 & 0 & 0 & 0 & 0 & 0 \\
        0 & \sqrt{\mu} R \sin \Theta & 0 & 0 & 0 & 0 & 0 & 0 \\
        0 & 0 & \sqrt{\mu} R & 0 & 0 & 0 & 0 & 0 \\
        0 & 0 & 0 & \sqrt{I_1} \sin \vartheta & 0 & \sqrt{I_1} \cos \vartheta & 0 & 0 \\
        0 & 0 & 0 & 0 & \sqrt{I_1} & 0 & 0 & 0 \\
        0 & 0 & 0 & 0 & 0 & \sqrt{I_1} & 0 & 0 \\
        0 & 0 & 0 & 0 & 0 & 0 & \sqrt{I_2} \sin \chi & 0 \\
        0 & 0 & 0 & 0 & 0 & 0 & 0 & \sqrt{I_2}
    \end{bmatrix} \mf{x}.
\end{gather}
The change of variables induces a Jacobian factor that can be obtained as a determinant of the $\bbT^\top$ matrix
\begin{gather}
    \Bigg\vert \frac{\partial \lsq \mf{q}, \mf{x} \rsq}{\partial \lsq \mf{q}, \mf{p} \rsq} \Bigg\vert = \text{det} \lb \bbT^\top \rb = \mu^{3/2} I_1^{3/2} I_2 R^2 \sin \Theta \sin \vartheta \sin \chi.
\end{gather}
The distribution $\rho_0$ in variables $\mf{x}$ splits into a product of independent Gaussian distributions
\begin{gather}
    \rho_0 \lb \mf{q}, \mf{x} \rb \sim R^2 \sin \Theta \sin \vartheta \sin \chi \prod_{k = 1}^{8} \exp \lb -\frac{1}{2} \beta x_k^2 \rb.  
\end{gather}
It follows that the vector components $x_k$ are distributed normally $x_k \sim \mathcal{N}(0, 1)$, cosines of each of the angles $\lc \Theta, \vartheta, \chi \rc$ are distributed uniformly in the range of $\lsq -1, 1 \rsq$, and the angles $\lc \Phi, \varphi, \psi, \eta \rc$ are distributed uniformly in the range of $\lsq 0, 2 \pi \rsq$. The samples of variable $R$ can be generated through selecting $R^3$ uniformly on $\lsq R^3_\text{min}, R^3_\text{max} \rsq$.

The sampling from the full Boltzmann distribution
\begin{gather}
    \rho \lb \mf{q}, \mf{p} \rb = \exp \lb -\frac{H}{k_\text{B} T} \rb
\end{gather}
is performed by applying the rejection criterion to the samples generated from the $\rho_0$-density (see details in Section IV.B of \citet{Chistikov2021}).  

\bibliography{biblio} 

\begin{thebibliography}{}
\expandafter\ifx\csname natexlab\endcsname\relax\def\natexlab#1{#1}\fi
\providecommand{\url}[1]{\href{#1}{#1}}
\providecommand{\dodoi}[1]{doi:~\href{http://doi.org/#1}{\nolinkurl{#1}}}
\providecommand{\doeprint}[1]{\href{http://ascl.net/#1}{\nolinkurl{http://ascl.net/#1}}}
\providecommand{\doarXiv}[1]{\href{https://arxiv.org/abs/#1}{\nolinkurl{https://arxiv.org/abs/#1}}}

\bibitem[{{Anderson} \& {Samuelson}(2011)}]{Anderson2011}
{Anderson}, C.~M., \& {Samuelson}, R.~E. 2011, Icarus, 212, 762,
  \dodoi{10.1016/j.icarus.2011.01.024}

\bibitem[{{Avice} {et~al.}(2018){Avice}, {Marty}, {Burgess}, {Hofmann},
  {Philippot}, {Zahnle}, \& {Zakharov}}]{Avice2018}
{Avice}, G., {Marty}, B., {Burgess}, R., {et~al.} 2018, Geochimica et
  Cosmochimica Acta, 232, 82, \dodoi{10.1016/j.gca.2018.04.018}

\bibitem[{{Bader} \& {Berne}(1994)}]{Bader1994}
{Bader}, J.~S., \& {Berne}, B. 1994, The Journal of chemical physics, 100, 8359

\bibitem[{{Berens} {et~al.}(1981){Berens}, {White}, \& {Wilson}}]{Berens1981}
{Berens}, P.~H., {White}, S.~R., \& {Wilson}, K.~R. 1981, The Journal of
  Chemical Physics, 75, 515

\bibitem[{{B{\'e}zard} \& {Vinatier}(2020)}]{Bezard2020}
{B{\'e}zard}, B., \& {Vinatier}, S. 2020, Icarus, 344, 113261,
  \dodoi{10.1016/j.icarus.2019.03.038}

\bibitem[{{B{\'e}zard} {et~al.}(2018){B{\'e}zard}, {Vinatier}, \&
  {Achterberg}}]{Bezard2018}
{B{\'e}zard}, B., {Vinatier}, S., \& {Achterberg}, R.~K. 2018, Icarus, 302,
  437, \dodoi{10.1016/j.icarus.2017.11.034}

\bibitem[{{Birnbaum} {et~al.}(1993){Birnbaum}, {Borysow}, \&
  {Buechele}}]{Birnbaum1993}
{Birnbaum}, G., {Borysow}, A., \& {Buechele}, A. 1993, Journal of Chemical
  Physics, 99, 3234, \dodoi{10.1063/1.465132}

\bibitem[{{Blake} {et~al.}(2010){Blake}, {Chang}, \& {Lepland}}]{Blake2010}
{Blake}, R.~E., {Chang}, S.~J., \& {Lepland}, A. 2010, Nature, 464, 1029,
  \dodoi{10.1038/nature08952}

\bibitem[{{Borysow} \& {Frommhold}(1986{\natexlab{a}})}]{Borysow1986-H2-N2}
{Borysow}, A., \& {Frommhold}, L. 1986{\natexlab{a}}, Astrophysical Journal,
  303, 495, \dodoi{10.1086/164096}

\bibitem[{{Borysow} \& {Frommhold}(1986{\natexlab{b}})}]{Borysow1986-N2-N2}
---. 1986{\natexlab{b}}, Astrophysical Journal, 311, 1043,
  \dodoi{10.1086/164841}

\bibitem[{{Borysow} \& {Frommhold}(1986{\natexlab{c}})}]{Borysow1986-CH4-H2}
---. 1986{\natexlab{c}}, Astrophysical Journal, 304, 849,
  \dodoi{10.1086/164221}

\bibitem[{{Borysow} \& {Frommhold}(1987)}]{Borysow1987-CH4-CH4}
---. 1987, Astrophysical Journal, 318, 940, \dodoi{10.1086/165426}

\bibitem[{{Borysow} \& {Moraldi}(1994)}]{Borysow1994}
{Borysow}, A., \& {Moraldi}, M. 1994, Molecular Physics, 82, 1277,
  \dodoi{10.1080/00268979400100904}

\bibitem[{{Borysow} \& {Tang}(1993)}]{Borysow1993}
{Borysow}, A., \& {Tang}, C. 1993, Icarus, 105, 175,
  \dodoi{10.1006/icar.1993.1117}

\bibitem[{{Borysow} {et~al.}(1985){Borysow}, {Moraldi}, \&
  {Frommhold}}]{Borysow1985}
{Borysow}, J., {Moraldi}, M., \& {Frommhold}, L. 1985, Molecular Physics, 56,
  913, \dodoi{10.1080/00268978500102801}

\bibitem[{{Bussery-Honvault} \& {Hartmann}(2014)}]{Bussery-Honvalut2014}
{Bussery-Honvault}, B., \& {Hartmann}, J.-M. 2014, Journal of Chemical Physics,
  140, 054309, \dodoi{10.1063/1.4863636}

\bibitem[{{Catling}(2014)}]{Catling2014}
{Catling}, D.~C. 2014, in {Treatise on Geochemistry}, ed. H.~D. {Holland} \&
  K.~K. {Turekian}, Vol.~8 (Elsevier), {191--233}

\bibitem[{{Catling} \& {Zahnle}(2020)}]{Catling2020}
{Catling}, D.~C., \& {Zahnle}, K.~J. 2020, Science Advances, 6, eaax1420,
  \dodoi{10.1126/sciadv.aax1420}

\bibitem[{{Charnay} {et~al.}(2020){Charnay}, {Wolf}, {Marty}, \&
  {Forget}}]{Charnay2020}
{Charnay}, B., {Wolf}, E.~T., {Marty}, B., \& {Forget}, F. 2020, Space Science
  Reviews, 216, 90, \dodoi{10.1007/s11214-020-00711-9}

\bibitem[{{Chistikov} {et~al.}(2021){Chistikov}, {Finenko}, {Kalugina},
  {Lokshtanov}, {Petrov}, \& {Vigasin}}]{Chistikov2021}
{Chistikov}, D.~N., {Finenko}, A.~A., {Kalugina}, Y.~N., {et~al.} 2021, Journal
  of Chemical Physics, \dodoi{10.1063/5.0060779}

\bibitem[{{Chistikov} {et~al.}(2018){Chistikov}, {Finenko}, {Lokshtanov},
  {Petrov}, \& {Vigasin}}]{Chistikov2018}
{Chistikov}, D.~N., {Finenko}, A.~A., {Lokshtanov}, S.~E., {Petrov}, S.~V., \&
  {Vigasin}, A.~A. 2018, Journal of Chemical Physics, 149, 194304,
  \dodoi{10.1063/1.5054125}

\bibitem[{{Chistikov} {et~al.}(2019){Chistikov}, {Finenko}, {Lokshtanov},
  {Petrov}, \& {Vigasin}}]{Chistikov2019}
---. 2019, Journal of Chemical Physics, 151, 194106, \dodoi{10.1063/1.5125756}

\bibitem[{{Courtin} {et~al.}(1995){Courtin}, {Gautier}, \&
  {McKay}}]{Courtin1995}
{Courtin}, R., {Gautier}, D., \& {McKay}, C.~P. 1995, Icarus, 114, 144,
  \dodoi{10.1006/icar.1995.1050}

\bibitem[{{Dagg} {et~al.}(1986){Dagg}, {Anderson}, {Yan}, {Smith}, {Joslin}, \&
  {Read}}]{Dagg1986}
{Dagg}, I.~R., {Anderson}, A., {Yan}, S., {et~al.} 1986, Canadian Journal of
  Physics, 64, 1467, \dodoi{10.1139/p86-260}

\bibitem[{{de Kok} {et~al.}(2010){de Kok}, {Irwin}, \& {Teanby}}]{deKok2010}
{de Kok}, R., {Irwin}, P.~G.~J., \& {Teanby}, N.~A. 2010, Icarus, 209, 854,
  \dodoi{10.1016/j.icarus.2010.06.035}

\bibitem[{{Doose} {et~al.}(2016){Doose}, {Karkoschka}, {Tomasko}, \&
  {Anderson}}]{Doose2016}
{Doose}, L.~R., {Karkoschka}, E., {Tomasko}, M.~G., \& {Anderson}, C.~M. 2016,
  \icarus, 270, 355, \dodoi{10.1016/j.icarus.2015.09.039}

\bibitem[{{Egelstaff}(1962)}]{Egelstaff1962}
{Egelstaff}, P.~A. 1962, Advances in Physics, 11, 203

\bibitem[{{Finenko} {et~al.}(2021){Finenko}, {Chistikov}, {Kalugina}, {Conway},
  \& {Gordon}}]{Finenko2021}
{Finenko}, A.~A., {Chistikov}, D.~N., {Kalugina}, Y.~N., {Conway}, E.~K., \&
  {Gordon}, I.~E. 2021, Physical Chemistry Chemical Physics,
  \dodoi{10.1039/d1cp02161c}

\bibitem[{{Flasar} {et~al.}(2004){Flasar}, {Kunde}, {Abbas}, {Achterberg},
  {Ade}, {Barucci}, {B{\'e}zard}, {Bjoraker}, {Brasunas}, {Calcutt}, {Carlson},
  {C{\'e}sarsky}, {Conrath}, {Coradini}, {Courtin}, {Coustenis}, {Edberg},
  {Edgington}, {Ferrari}, {Fouchet}, {Gautier}, {Gierasch}, {Grossman},
  {Irwin}, {Jennings}, {Lellouch}, {Mamoutkine}, {Marten}, {Meyer}, {Nixon},
  {Orton}, {Owen}, {Pearl}, {Prang{\'e}}, {Raulin}, {Read}, {Romani},
  {Samuelson}, {Segura}, {Showalter}, {Simon-Miller}, {Smith}, {Spencer},
  {Spilker}, \& {Taylor}}]{Flasar2004}
{Flasar}, F.~M., {Kunde}, V.~G., {Abbas}, M.~M., {et~al.} 2004, Space Science
  Reviews, 115, 169, \dodoi{10.1007/s11214-004-1454-9}

\bibitem[{Frommhold(2006)}]{Frommhold2006}
Frommhold, L. 2006, Collision Induced Absorption in Gases (Cambridge University
  Press)

\bibitem[{{Fulchignoni} {et~al.}(2005){Fulchignoni}, {Ferri}, {Angrilli},
  {Ball}, {Bar-Nun}, {Barucci}, {Bettanini}, {Bianchini}, {Borucki},
  {Colombatti}, {Coradini}, {Coustenis}, {Debei}, {Falkner}, {Fanti},
  {Flamini}, {Gaborit}, {Grard}, {Hamelin}, {Harri}, {Hathi}, {Jernej},
  {Leese}, {Lehto}, {Lion Stoppato}, {L{\'o}pez-Moreno}, {M{\"a}kinen},
  {McDonnell}, {McKay}, {Molina-Cuberos}, {Neubauer}, {Pirronello}, {Rodrigo},
  {Saggin}, {Schwingenschuh}, {Seiff}, {Sim{\~o}es}, {Svedhem}, {Tokano},
  {Towner}, {Trautner}, {Withers}, \& {Zarnecki}}]{Fulchignoni2005}
{Fulchignoni}, M., {Ferri}, F., {Angrilli}, F., {et~al.} 2005, \nat, 438, 785,
  \dodoi{10.1038/nature04314}

\bibitem[{{Gillon} {et~al.}(2017){Gillon}, {Triaud}, {Demory}, {Jehin}, {Agol},
  {Deck}, {Lederer}, {de Wit}, {Burdanov}, {Ingalls}, {Bolmont}, {Leconte},
  {Raymond}, {Selsis}, {Turbet}, {Barkaoui}, {Burgasser}, {Burleigh}, {Carey},
  {Chaushev}, {Copperwheat}, {Delrez}, {Fernandes}, {Holdsworth}, {Kotze}, {Van
  Grootel}, {Almleaky}, {Benkhaldoun}, {Magain}, \& {Queloz}}]{Gillon2017}
{Gillon}, M., {Triaud}, A. H.~M.~J., {Demory}, B.-O., {et~al.} 2017, Nature,
  542, 456, \dodoi{10.1038/nature21360}

\bibitem[{{Goldstein} {et~al.}(2001){Goldstein}, {Poole}, \&
  {Safko}}]{Goldstein2001}
{Goldstein}, H., {Poole}, C.~P., \& {Safko}, J.~L. 2001, {Classical Mechanics}
  (Addison Wesley Publishing Company, San Francisco)

\bibitem[{{Gordon} {et~al.}(2021){Gordon}, {Rothman}, {Hargreaves}, {Hashemi},
  {Karlovets}, {Skinner}, {Conway}, {Hill}, {Kochanov}, {Tan}, {Wcis{\l}o},
  {Finenko}, {Nelson}, {Bernath}, {Birk}, {Boudon}, {Campargue}, {Chance},
  {Coustenis}, {Drouin}, {Flaud}, {Gamache}, {Hodges}, {Jacquemart}, {Mlawer},
  {Nikitin}, {Perevalov}, {Rotger}, {Tennyson}, {Toon}, {Tran}, {Tyuterev},
  {Adkins}, {Baker}, {Barbe}, {Can{\`e}}, {Cs{\'a}sz{\'a}r}, {Dudaryonok},
  {Egorov}, {Fleisher}, {Fleurbaey}, {Foltynowicz}, {Furtenbacher}, {Harrison},
  {Hartmann}, {Horneman}, {Huang}, {Karman}, {Karns}, {Kassi}, {Kleiner},
  {Kofman}, {Kwabia--Tchana}, {Lavrentieva}, {Lee}, {Long}, {Lukashevskaya},
  {Lyulin}, {Makhnev}, {Matt}, {Massie}, {Melosso}, {Mikhailenko}, {Mondelain},
  {M\"{u}ller}, {Naumenko}, {Perrin}, {Polyansky}, {Raddaoui}, {Raston},
  {Reed}, {Rey}, {Richard}, {T{\'{o}}bi{\'{a}}s}, {Sadiek}, {Schwenke},
  {Starikova}, {Sung}, {Tamassia}, {Tashkun}, {Vander Auwera}, {Vasilenko},
  {Vigasin}, {Villanueva}, {Vispoel}, {Wagner}, {Yachmenev}, \&
  {Yurchenko}}]{HITRAN2020}
{Gordon}, I.~E., {Rothman}, L.~S., {Hargreaves}, R.~J., {et~al.} 2021, Journal
  of Quantitative Spectroscopy and Radiative Transfer, 107949,
  \dodoi{10.1016/j.jqsrt.2021.107949}

\bibitem[{{Gruszka} \& {Borysow}(1997)}]{Gruszka1997}
{Gruszka}, M., \& {Borysow}, A. 1997, Icarus, 129, 172,
  \dodoi{10.1006/icar.1997.5773}

\bibitem[{{Hartmann} {et~al.}(2011){Hartmann}, {Boulet}, \&
  {Jacquemart}}]{Hartmann2011-II}
{Hartmann}, J.~M., {Boulet}, C., \& {Jacquemart}, D. 2011, Journal of Chemical
  Physics, 134, 094316, \dodoi{10.1063/1.3557681}

\bibitem[{Hindmarsh {et~al.}(2005)Hindmarsh, Brown, Grant, Lee, Serban,
  Shumaker, \& Woodward}]{SUNDIALS}
Hindmarsh, A.~C., Brown, P.~N., Grant, K.~E., {et~al.} 2005, ACM Trans. Math.
  Softw., 31, 363–396, \dodoi{10.1145/1089014.1089020}

\bibitem[{{Hren} {et~al.}(2009){Hren}, {Tice}, \& {Chamberlain}}]{Hren2009}
{Hren}, M.~T., {Tice}, M.~M., \& {Chamberlain}, C.~P. 2009, Nature, 462, 205,
  \dodoi{10.1038/nature08518}

\bibitem[{{Hunten} {et~al.}(1984){Hunten}, {Tomasko}, {Flasar}, {Samuelson},
  {Strobe}, \& {Stevenson}}]{Hunten1984}
{Hunten}, D.~M., {Tomasko}, M.~G., {Flasar}, F.~M., {et~al.} 1984, in Saturn,
  ed. T.~{Gehrels} \& N.~{Shapley Matthews} (Univ. Arizona Press, Tucson),
  {671--759}

\bibitem[{{Jennings} {et~al.}(2017){Jennings}, {Flasar}, {Kunde}, {Nixon},
  {Segura}, {Romani}, {Gorius}, {Albright}, {Brasunas}, {Carlson},
  {Mamoutkine}, {Guandique}, {Kaelberer}, {Aslam}, {Achterberg}, {Bjoraker},
  {Anderson}, {Cottini}, {Pearl}, {Smith}, {Hesman}, {Barney}, {Calcutt},
  {Vellacott}, {Spilker}, {Edgington}, {Brooks}, {Ade}, {Schinder},
  {Coustenis}, {Courtin}, {Michel}, {Fettig}, {Pilorz}, \&
  {Ferrari}}]{Jennings2017}
{Jennings}, D.~E., {Flasar}, F.~M., {Kunde}, V.~G., {et~al.} 2017, Applied
  Optics, 56, 5274, \dodoi{10.1364/ao.56.005274}

\bibitem[{{Karman} {et~al.}(2015){Karman}, {Miliordos}, {Hunt}, {Groenenboom},
  \& {van der Avoird}}]{Karman2015}
{Karman}, T., {Miliordos}, E., {Hunt}, K. L.~C., {Groenenboom}, G.~C., \& {van
  der Avoird}, A. 2015, Journal of Chemical Physics, 142, 084306,
  \dodoi{10.1063/1.4907917}

\bibitem[{{Karman} {et~al.}(2019){Karman}, {Gordon}, {van der Avoird},
  {Baranov}, {Boulet}, {Drouin}, {Groenenboom}, {Gustafsson}, {Hartmann},
  {Kurucz}, {Rothman}, {Sun}, {Sung}, {Thalman}, {Tran}, {Wishnow},
  {Wordsworth}, {Vigasin}, {Volkamer}, \& {van der Zande}}]{Karman2019}
{Karman}, T., {Gordon}, I.~E., {van der Avoird}, A., {et~al.} 2019, Icarus,
  328, 160, \dodoi{10.1016/j.icarus.2019.02.034}

\bibitem[{{Krissansen-Totton} {et~al.}(2018{\natexlab{a}}){Krissansen-Totton},
  {Arney}, \& {Catling}}]{Krissansen-Totton2018-II}
{Krissansen-Totton}, J., {Arney}, G.~N., \& {Catling}, D.~C.
  2018{\natexlab{a}}, Proceedings of the National Academy of Science, 115,
  4105, \dodoi{10.1073/pnas.1721296115}

\bibitem[{{Krissansen-Totton} {et~al.}(2018{\natexlab{b}}){Krissansen-Totton},
  {Olson}, \& {Catling}}]{Krissansen-Totton2018-I}
{Krissansen-Totton}, J., {Olson}, S., \& {Catling}, D.~C. 2018{\natexlab{b}},
  Science Advances, 4, eaao5747, \dodoi{10.1126/sciadv.aao5747}

\bibitem[{{Lax}(2007)}]{Lax2007}
{Lax}, P.~D. 2007, Linear Algebra and Its Applications, 2nd edn. (John Wiley \&
  Sons Inc, Hoboken, New Jersey)

\bibitem[{{Lepage}(1978)}]{Lepage1978}
{Lepage}, G.~P. 1978, Journal of Computational Physics, 27, 192,
  \dodoi{10.1016/0021-9991(78)90004-9}

\bibitem[{{Lichtenberg}(2021)}]{Lichtenberg2021}
{Lichtenberg}, T. 2021, Astrophysical Journal Letters, 914, L4,
  \dodoi{10.3847/2041-8213/ac0146}

\bibitem[{{Lorenz} {et~al.}(1997){Lorenz}, {McKay}, \& {Lunine}}]{Lorenz1997}
{Lorenz}, R.~D., {McKay}, C.~P., \& {Lunine}, J.~I. 1997, Science, 275, 642,
  \dodoi{10.1126/science.275.5300.642}

\bibitem[{{Niemann} {et~al.}(2005){Niemann}, {Atreya}, {Bauer}, {Carignan},
  {Demick}, {Frost}, {Gautier}, {Haberman}, {Harpold}, {Hunten}, {Israel},
  {Lunine}, {Kasprzak}, {Owen}, {Paulkovich}, {Raulin}, {Raaen}, \&
  {Way}}]{Niemann2005}
{Niemann}, H.~B., {Atreya}, S.~K., {Bauer}, S.~J., {et~al.} 2005, Nature, 438,
  779, \dodoi{10.1038/nature04122}

\bibitem[{{Niemann} {et~al.}(2010){Niemann}, {Atreya}, {Demick}, {Gautier},
  {Haberman}, {Harpold}, {Kasprzak}, {Lunine}, {Owen}, \&
  {Raulin}}]{Niemann2010}
{Niemann}, H.~B., {Atreya}, S.~K., {Demick}, J.~E., {et~al.} 2010, Journal of
  Geophysical Research (Planets), 115, E12006, \dodoi{10.1029/2010JE003659}

\bibitem[{{Odintsova} {et~al.}(2021){Odintsova}, {Serov}, {Balashov},
  {Koshelev}, {Koroleva}, {Simonova}, {Tretyakov}, {Filippov}, {Chistikov},
  {Finenko}, {Lokshtanov}, {Petrov}, \& {Vigasin}}]{Odintsova2021}
{Odintsova}, T.~A., {Serov}, E.~A., {Balashov}, A.~A., {et~al.} 2021, Journal
  of Quantitative Spectroscopy and Radiative Transfer, 258, 107400,
  \dodoi{10.1016/j.jqsrt.2020.107400}

\bibitem[{{Oparin} {et~al.}(2017){Oparin}, {Filippov}, {Grigoriev}, \&
  {Kouzov}}]{Oparin2017}
{Oparin}, D.~V., {Filippov}, N.~N., {Grigoriev}, I.~M., \& {Kouzov}, A.~P.
  2017, Journal of Quantitative Spectroscopy and Radiative Transfer, 196, 87,
  \dodoi{10.1016/j.jqsrt.2017.04.002}

\bibitem[{{Pavlov} {et~al.}(2001){Pavlov}, {Brown}, \& {Kasting}}]{Pavlov2001}
{Pavlov}, A.~A., {Brown}, L.~L., \& {Kasting}, J.~F. 2001, Journal of
  Geophysical Research, 106, 23267, \dodoi{10.1029/2000JE001448}

\bibitem[{{Pavlov} {et~al.}(2000){Pavlov}, {Kasting}, {Brown}, {Rages}, \&
  {Freedman}}]{Pavlov2000}
{Pavlov}, A.~A., {Kasting}, J.~F., {Brown}, L.~L., {Rages}, K.~A., \&
  {Freedman}, R. 2000, Journal of Geophysical Research, 105, 11981,
  \dodoi{10.1029/1999JE001134}

\bibitem[{{Samuelson} {et~al.}(1997){Samuelson}, {Nath}, \&
  {Borysow}}]{Samuelson1997}
{Samuelson}, R.~E., {Nath}, N.~R., \& {Borysow}, A. 1997, Planetary and Space
  Science, 45, 959, \dodoi{https://doi.org/10.1016/S0032-0633(97)00090-1}

\bibitem[{{Schofield}(1960)}]{Schofield1960}
{Schofield}, P. 1960, Physical Review Letters, 4, 239,
  \dodoi{10.1103/PhysRevLett.4.239}

\bibitem[{{Sung} {et~al.}(2016){Sung}, {Wishnow}, {Venkataraman}, {Brown},
  {Ozier}, {Benner}, {Crawford}, {Mantz}, \& {Smith}}]{Sung2016}
{Sung}, K., {Wishnow}, E., {Venkataraman}, M., {et~al.} 2016, in AAS/Division
  for Planetary Sciences Meeting Abstracts, Vol.~48, AAS/Division for Planetary
  Sciences Meeting Abstracts \#48, 424.11

\bibitem[{{Tomasko} {et~al.}(2008){Tomasko}, {B{\'e}zard}, {Doose}, {Engel},
  {Karkoschka}, \& {Vinatier}}]{Tomasko2008}
{Tomasko}, M.~G., {B{\'e}zard}, B., {Doose}, L., {et~al.} 2008, Planetary and
  Space Science, 56, 648, \dodoi{10.1016/j.pss.2007.10.012}

\bibitem[{{Turbet} {et~al.}(2018){Turbet}, {Bolmont}, {Leconte}, {Forget},
  {Selsis}, {Tobie}, {Caldas}, {Naar}, \& {Gillon}}]{Turbet2018}
{Turbet}, M., {Bolmont}, E., {Leconte}, J., {et~al.} 2018, Astronomy \&
  Astrophysics, 612, A86, \dodoi{10.1051/0004-6361/201731620}

\bibitem[{{van Kranendonk} \& {Gass}(1973)}]{Kranendonk1973}
{van Kranendonk}, J., \& {Gass}, D.~M. 1973, Canadian Journal of Physics, 51,
  2428, \dodoi{10.1139/p73-317}

\bibitem[{{Wordsworth} {et~al.}(2017){Wordsworth}, {Kalugina}, {Lokshtanov},
  {Vigasin}, {Ehlmann}, {Head}, {Sanders}, \& {Wang}}]{Wordsworth2017}
{Wordsworth}, R., {Kalugina}, Y., {Lokshtanov}, S., {et~al.} 2017, Geophysical
  Research Letters, 44, 665, \dodoi{10.1002/2016GL071766}

\bibitem[{{Zahnle} {et~al.}(2006){Zahnle}, {Claire}, \& {Catling}}]{Zahnle2006}
{Zahnle}, K.~J., {Claire}, M.~W., \& {Catling}, D.~C. 2006, Geobiology, 4, 271,
  \dodoi{https://doi.org/10.1111/j.1472-4669.2006.00085.x}

\bibitem[{{Zahnle} {et~al.}(2019){Zahnle}, {Gacesa}, \& {Catling}}]{Zahnle2019}
{Zahnle}, K.~J., {Gacesa}, M., \& {Catling}, D.~C. 2019, Geochimica et
  Cosmochimica Acta, 244, 56, \dodoi{10.1016/j.gca.2018.09.017}

\end{thebibliography}
\bibliographystyle{aasjournal}

\end{document}